\documentclass[sigconf]{acmart}
\pdfoutput=1
\usepackage{epsfig}
\usepackage{endnotes}
\usepackage{verbatim}
\usepackage{gensymb}
\usepackage{textcomp}
\usepackage{subcaption}
\usepackage{graphics}
\usepackage{graphicx}

\usepackage{multirow}
\usepackage{amsmath}

\usepackage{listings}
\usepackage{verbatim}
\usepackage{amssymb}

\usepackage{amsthm}
\usepackage{alltt}
\usepackage{multirow}
\usepackage{graphicx}
\usepackage{amsfonts}
\usepackage{wrapfig}
\usepackage{boxedminipage}

\usepackage{rotating}
\usepackage{color}
\usepackage{xspace}
\usepackage{setspace}
\usepackage{url}
\usepackage{lscape}
\usepackage{syntax}
\usepackage{proof}
\usepackage{array}
\usepackage{caption}
\usepackage{epsfig}
\usepackage{endnotes}
\usepackage{bbm}
\usepackage{mdframed}
\usepackage{float}
\usepackage{bm}

\usepackage{amssymb}
\usepackage{algorithm}
\usepackage{algpseudocode}
\usepackage{soul}
\usepackage{xifthen}
\usepackage{mathtools}
\usepackage{stackrel}
\usepackage[outdir=./]{epstopdf}

\graphicspath{{./Figures/}}
\newcommand{\name}{\textsc{Zeus}\xspace}

\definecolor{linkcol}{rgb}{0,0,1}
\definecolor{citecol}{rgb}{0,0.5,0}
\definecolor{urlcol}{rgb}{0.3,0,0}

\begin{document}
\sloppy

\copyrightyear{2017}
\acmYear{2017}
\setcopyright{acmcopyright}
\acmConference{CCS '17}{October 30-November 3, 2017}{Dallas, TX, USA}
\acmPrice{15.00}
\acmDOI{10.1145/3133956.3134081} 
\acmISBN{978-1-4503-4946-8/17/10}

\fancyhead{}
\settopmatter{printacmref=false, printfolios=false}

\title{Watch Me, but Don't Touch Me! Contactless Control Flow Monitoring via Electromagnetic Emanations} 

% \author{
%   Yi Han, Sriharsha Etigowni, Hua Liu, Saman Zonouz, Athina Petropulu\\
%   Electrical and Computer Engineering, Rutgers University\\
%   \textit{\{yh482, hl678, sriharsha.etigowni, saman.zonouz, athinap\}@rutgers.edu}
% }

% \author{Yi Han, Sriharsha Etigowni, Hua Liu, Saman Zonouz, Athina Petropulu}
% \affiliation{Electrical and Computer Engineering, Rutgers University}
%   \textit{\{yh482, hl678, sriharsha.etigowni, saman.zonouz, athinap\}@rutgers.edu}
% }

\author{Yi Han}
\affiliation{Rutgers University}
\email{yi.han@rutgers.edu}
\author{Sriharsha Etigowni}
\affiliation{Rutgers University}
\email{sriharsha.etigowni@rutgers.edu}
\author{Hua Liu}
\affiliation{Rutgers University}
\email{hl678@rutgers.edu}
\author{Saman Zonouz}
\affiliation{Rutgers University}
\email{saman.zonouz@rutgers.edu}
\author{Athina Petropulu}
\affiliation{Rutgers University}
\email{athinap@rutgers.edu} 

% \author{Yi Han}
% \email{yi.han@rutgers.edu}
% \author{Sriharsha Etigowni}
% \email{sriharsha.etigowni@rutgers.edu}
% \author{Hua Liu}
% \email{hl678@rutgers.edu}
% \author{Saman Zonouz}
% \email{saman.zonouz@rutgers.edu}
% \author{Athina Petropulu}
% \affiliation{Rutgers University}
% \email{athinap@rutgers.edu}

\begin{abstract}

Trustworthy operation of industrial control systems depends on secure and real-time code execution on the embedded programmable logic controllers (PLCs). The controllers monitor and control the critical infrastructures, such as electric power grids and health-care platforms, and continuously report back the system status to human operators. %
%\footnote{``Don't Touch Me'' is the name of a classic country music song originally recorded and made famous by Jeannie Seely in 1966.}
We present \name, a contactless embedded controller security monitor to ensure its execution control flow integrity. %
\name leverages the electromagnetic emission by the PLC circuitry during the execution of the controller programs. \name's contactless execution tracking enables non-intrusive monitoring of security-critical controllers with tight real-time constraints. Those devices often cannot tolerate the cost and performance overhead that comes with additional traditional hardware or software monitoring modules. Furthermore, \name provides an air-gap between the monitor (trusted computing base) and the target (potentially compromised) PLC. This eliminates the possibility of the monitor infection by the same attack vectors. 

\name monitors for control flow integrity of the PLC program execution. \name monitors the communications between the human-machine interface and the PLC, and captures the control logic binary uploads to the PLC. \name exercises its feasible execution paths, and fingerprints their emissions using an external electromagnetic sensor. \name trains a neural network for legitimate PLC executions, and uses it at runtime to identify the control flow based on PLC's electromagnetic emissions.  We implemented \name on a commercial Allen Bradley PLC, which is widely used in industry, and evaluated it on real-world control program executions. \name was able to distinguish between different legitimate and malicious executions  with 98.9\% accuracy and with zero overhead on PLC execution by design.

\end{abstract}

% <concept_id>10002978.10003022.10003028</concept_id>
% <concept_desc>Security and privacy~Domain-specific security and privacy architectures</concept_desc>

% \ccsdesc[500]{Security and privacy~Domain-specific security and privacy architectures}


\maketitle

\section{Introduction}
\label{sec:introduction}

%general problem
Industrial control systems (ICS) are fundamental parts of modern society as they control and monitor critical infrastructures such as electricity grids, health-care, chemical production, oil and gas refinery, transportation and water treatment. Due to their importance and large attack surfaces, they are becoming attractive targets for malicious penetrations leading to catastrophic failures with substantive impact~\cite{slay-07,poland-train} including the recent BlackEnergy worm against Ukranian electricity grid~\cite{blackenergy}. Recently, the Stuxnet malware uploaded malicious code to programmable logic controllers (PLCs), and physically damaged 20\% of Iranian PLC-controlled centrifuges~\cite{stuxnet-2010}. The discovery of Duqu~\cite{duqu-2011} and Havex~\cite{rrushiquantitative} show that such attacks are not isolated cases as they infected ICS in more than eight countries. Some of these vulnerable controllers are Internet-connected~\cite{enisa} and exposed by computer search engines like Shodan~\cite{bodenheim2014evaluation}. There have been an increasing number of reports on malicious attempts of PLC port scanning, automated PLC malware generation, modifying control system-specific protocols and access to system diagnostics~\cite{buza2014cryplh, mclaughlin2012sabot}.  Nevertheless, the ICS market is expected to grow to $\$10.33$ billion by 2018~\cite{TechNavio}.

%what others have done
There has been an increasing number of past works on embedded systems and PLC protection. Offline formal control logic analysis have been investigated by solutions such as TSV~\cite{mclaughlin2014tsv}, through symbolic execution and model checking mechanisms. Solutions such as WeaselBoard~\cite{mulder2013weaselboard} and CPAC~\cite{cpac} perform runtime PLC execution monitoring using control logic and firmware-level reference monitor implementations. Most related to our paper, there have been attack and defense solutions that employ side-channel analyses to either disclose secret information (e.g., cryptographic keys~\cite{genkin2016ecdsa}), or detect anomalous misbehavior (e.g., execution tracking~\cite{liu2016code}).  Side channel-based attacks require selective monitoring of only  execution points of interest, such as the encryption subroutines. On the other hand, side channel-based defenses have to monitor throughout the execution looking for anomalies.

%our work (high level)
In this paper, we present \name, a contactless PLC control flow integrity monitor that monitors the program execution by analyzing the PLC's runtime electromagnetic (EM) emanation side channel. Given a PLC controller program, \name profiles the PLC's electromagnetic emanation during the execution of feasible paths of the legitimate program. \name pre-processes the signal traces and uses them offline to train a neural network model. The model is later used during the runtime operation to either determine the fine-grained control flow of the execution based on the real-time EM emanations or declare unknown (malicious) code execution. %

Contactless monitoring enables \name to ensure security of crucial controllers in mission-critical applications with tight real-time constraints. The operators are often very reluctant to instrument those controllers' software stack with additional security probes that cause performance overhead and hurt the underlying real-time guarantees. Additionally, from security viewpoint, contactless monitoring keeps \name away from the attack vectors that target the controllers because of the introduced air gap between the monitor and the victim controller. Other side channel-based techniques such as power signal analyses draw and monitor current from controllers' circuitry. In contrast, \name is completely non-intrusive and passive; it does not require any instrumentation of the controller and does not affect its electronics. 

%our work (technical)
\name monitors all the network links bound to the PLC, and captures the control logic uploads by the human-machine interface (HMI) servers that are sent for execution on the PLC. \name exercises various code segments of the binary while capturing the electromagnetic emanations. \name profiles the control logic on the PLC with deactivated output modules to ensure the underlying physical process (actuators) are not affected during the training phase. The training is implemented in two stages. First, \name executes control logic symbolically\footnote{Complete symbolic execution of embedded PLC control logic programs is often feasible as they are mostly not branch-heavy in practice.} and removes infeasible paths. Through counterexample guided inductive synthesis~\cite{solar2008program}, \name generates different test inputs for each execution path, and trains a neural network based on the corresponding electromagnetic emanation. 
The trained neural network allows \name to detect the execution of an illegitimate control flow and/or malicious code injection. %Yi's comment out \textcolor{brown}{discussion about HMM and gadget and data flow}\textcolor{red}{\name also implements an on-demand more fine-grained execution monitoring by modeling the control flow graph as a hidden Markov model (HMM), and training its corresponding parameters. The trained HMM enables \name to localize the illegitimate execution path down to  basic block-level granularity. As a part of the offline analysis, \name synthesizes a set of small code gadgets with reconfigurable execution lengths, which can be injected into the target control logic program as electromagnetic markers. The gadgets serve two purposes: \textit{i)} they increase the ``visibility'' of a selected set of program points with weak electromagnetic emanations; and \textit{ii)} they encode and make a predefined set of program variable values visible to the external sensor. The encoding schemes varies over time using keyed hash function chain to prevent malicious eavesdroppers from reverse engineering the underlying mappings between the electromagnetic signals and the control/data flows.  By exploiting the typical time gaps between the cyclic scan-cycles due to the PLC-specific synchronous execution paradigm, the injected gadgets do not result in performance overhead.}

%contributions
The contributions of this paper are as follows: 
\begin{itemize}
	\item We present a new execution control flow tracking solution for embedded PLC controllers that enables security monitoring with air-gapped electromagnetic sensors.
    \item We develop an online signal processing framework to analyze the electromagnetic signals and extract minimal feature set necessary for execution integrity monitoring.
    \item We  evaluated \name using an inexpensive sensor against widely-used control programs, e.g., proportional-integral-derivative (PID) controllers, on commercial Allen Bradley PLC devices (most popular in North  America) with ARM Cortex-M3 processors. \name detects malicious code injections with $98.9\%$ accuracy in real-world settings.
    \end{itemize}

%implementations
%developed a linear PLC binary decompiler and program analysis modules, and

%organization
\paragraph{\textbf{Overview and Organization.}} In \autoref{sec:threat-model}, we explain our assumptions about the adversaries and their capabilities. In \autoref{sec:background}, we provide a brief background on programmable logic controllers and their typical configurations as well as neural networks that \name employs for program behavioral modeling. In \autoref{sec:emanation-analysis}, we discuss about the electromagnetic signals emitted by the PLCs and how they characterize the program execution. We discuss how \name generates training data points for program behavioral modeling and transforms the signal traces into spectrum sequences. In \autoref{sec:control-flow}, we present our fine-grained emanation analysis model, where a neural network model of the legitimate program control flows is constructed and trained using electromagnetic emanation signals. In \autoref{sec:evaluation}, we present our empirical evaluations of \name's various components on ten real-world PLC programs and attack scenarios similar to Stuxnet~\cite{stuxnet-2010}.  In \autoref{sec:related-work}, we review the recent most related work in the literature, and finally, we conclude the paper in \autoref{sec:conclusions}.

\section{Threat model}
\label{sec:threat-model}

One of the most prominent security failure causes in control systems using PLCs is the failure to guard PLCs against remote programming~\cite{wueest2014targeted}. PLC programmer machines are most often based on commodity operating systems, and often lag security update releases by months~\cite{ics-malware}. In the following, we state the assumptions made on the security measures that must be successfully in place for \name to function correctly. 

We assume there is some trusted path from \name to system operators to alert them of any malicious executions. Unlike software-based solutions, \name's contactless monitoring enables secure monitoring even if the software stack below the PLC's control logic (e.g., firmware or operating system) is compromised. The PLC-bound network link used to transfer the control logic programs for execution is assumed to be trusted.  This allows \name to obtain a legitimate copy of the control logic to compare with the runtime PLC executions for control flow integrity checks. \name does not assume source code availability and works with binaries. 

\name defends against \textit{control channel} attacks (e.g., Stuxnet~\cite{stuxnet-2010}), where the adversaries upload arbitrary and potentially malicious control logic on the PLC for execution. More specifically, the types of control logic attacks that \name can protect against are \textit{i)} modified control logic such as injection, removal, and replacement of code segments in the legitimate control logic program; and \textit{ii)} hijacked control flow of the legitimate control logic execution through network exploits (e.g., code reuse attacks\footnote{Protection against control flow and code reuse attacks are simpler in PLCs compared to conventional computers, because the PLCs' restrictive and more primitive programming languages (e.g., type safe and no indirect call sites) allows deterministic modeling of the legitimate control flows.} such as return-oriented programming). \name does not defend against \textit{sensor channel} attacks, where sensor data is forged by compromised sensors. In such a case, the control logic may behave exactly as intended, but on false sensor data~\cite{liu2011false}.
Additionally, Zeus itself may be attacked by external signal jammers leading to false positives. However, this would not affect the integrity of the control logic execution on the PLC. % This will constantly trigger Zeus to report anomaly even though the control logic program is not attacked. However, in this paper we focus on attacks that is against the PLC, such jamming attacks won't be able to let malicious codes running on the PLC bypass our detection system.

\section{Background}
\label{sec:background}

\paragraph{\textbf{Programmable Logic Controllers.}} Programmable logic controllers are multiple-input-multiple-output computers. They have input and output modules to interact with the physical world (plant) to monitor and control critical infrastructures such as manufacturing, robotics, and avionics. The PLC's input modules are connected to sensors within the plant and receive measurements about the plant's status continuously. The PLC's output modules are connected to plant actuators and convey the commands to control it. The PLC converts sensor readings into digital values, process the readings with the built-in computing unit, and forward the outputs to the physical world. The logical behavior of PLCs (i.e., the processing of the input data) is programmable. %
The control logic programs are developed by the control system operators on human-machine-interface (HMI) servers that are connected to the PLCs through network links. Once developed, the control logic is compiled and sent to the PLC for execution. The program is executed repeatedly in fixed intervals, called scan cycles. During each scan cycle, the control logic program reads input values from memory and stores the output values to memory. The PLC firmware is responsible for the interchange of these updated values to and from the PLC's input/output ports to interface the physical world. The firmware also implements the reporting mechanisms such as the LED display on the device and real-time data transfers to the HMI about the plant's current status.

\begin{figure}[tp]
  \centering
  \includegraphics[scale=0.3]{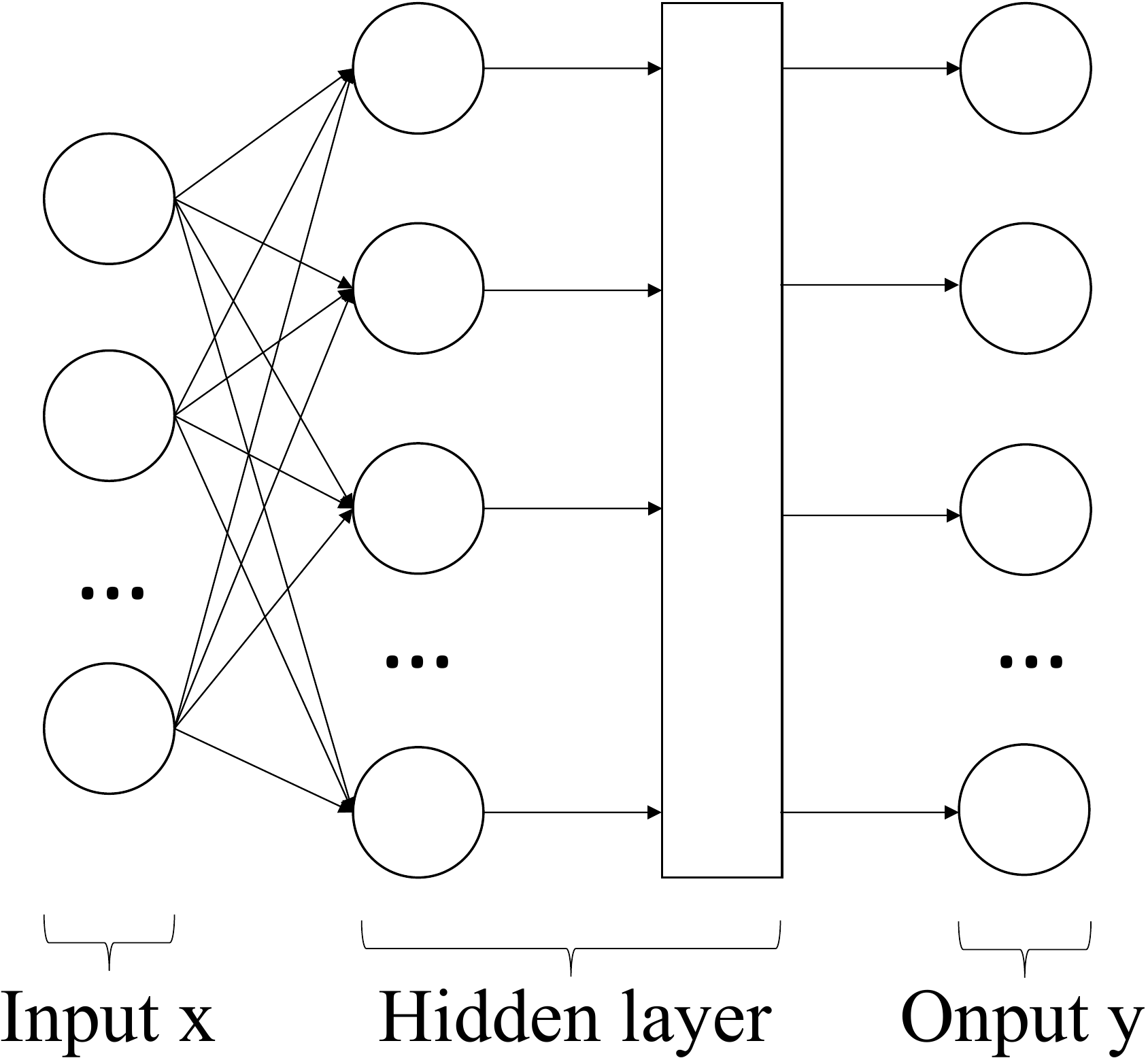}
  \caption{A basic neural network unit architecture.}
   \label{fig:neural}
\end{figure}

\paragraph{\textbf{Deep Neural Networks.}} Neural network is a class of supervised learning models that tries to learn the complex nonlinear mapping between input data and their targets (e.g. class labels). A basic neural network unit architecture consists of a linear mapping followed by an activation:

\begin{equation}
y=\sigma(Wx+b),
\label{eq:neural}
\end{equation}
where $x$ is the input feature vector, and $W$ represents the weight matrix. $\sigma$ is the activation function. It is a nonlinear function that models the complex relation between input $x$ and output $y$. Common activation functions include sigmoid~\cite{iliev2017approximation}, rectified linear unit (ReLU)~\cite{zhang2017beyond}, $\tanh$, etc. \autoref{fig:neural} shows a graphical illustration of \autoref{eq:neural}. The edges between the Input layer and the hidden layer represents the weights $W, b$. Since each node in the input layer is fully connected with all nodes in the hidden layer, such unit is also called a dense layer. 

A neural network can go large, which is increasing number of nodes in the hidden layer, or go deep, which is stacking multiple network units together (increasing number of hidden layers), such that more complex nonlinear mappings between data and targets can be learned. All forms of artificial neural networks essentially follow the aforementioned basic architecture. \name utilizes recurrent neural network (RNN)~\cite{ravuri2016comparative} to model the execution behavior of PLC programs (\autoref{sec:control-flow}). 

For training, the data samples, each with a corresponding label, are fed to the network's input layer. The network is trained to learn discriminative features from samples by itself. This completely data driven approach, compared to traditional hand crafted feature extraction methods, leads to much simpler-to-use and more reliable outcomes in practice. In our experiments (\autoref{sec:evaluation}), we empirically show that RNNs overcome traditional techniques such as hidden Markov models (HMMs) in terms of PLC execution monitoring accuracy and performance. 

Neural networks can be trained in an iterative manner using the gradient descent algorithm~\cite{yuan2016convergence}. At each iteration, all input data are passed through the network. The output are compared with their corresponding targets $t$. A loss function $l$ is defined between the network output $y$ and the expected target on each data sample:

\begin{equation}
l_i = loss(y_i, t_i),
\end{equation}
The loss function measures the difference between the current and target outputs. \name uses mean square error (MSE) as the loss function. The total loss is the sum over individual losses of all the data samples:

\begin{equation}
l=\sum_{i=1}^{N} l_i,
\end{equation}
where $N$ represents the total number of data samples. Computing the total loss is called the forward pass. To update weights or parameters of the network, partial derivatives of the total loss with respect to all weights are calculated to identify their maximal descending direction using back propagation. All weights are updated accordingly (the backward pass). Forward and backward passes are repeated iteratively until the values converge. The resulting network is able to produce outputs close to the expected targets.

\section{PLC Program Emanation Analysis} 
\label{sec:emanation-analysis}

\begin{figure*}[tp]
  \centering
  \includegraphics[width=\textwidth, ]{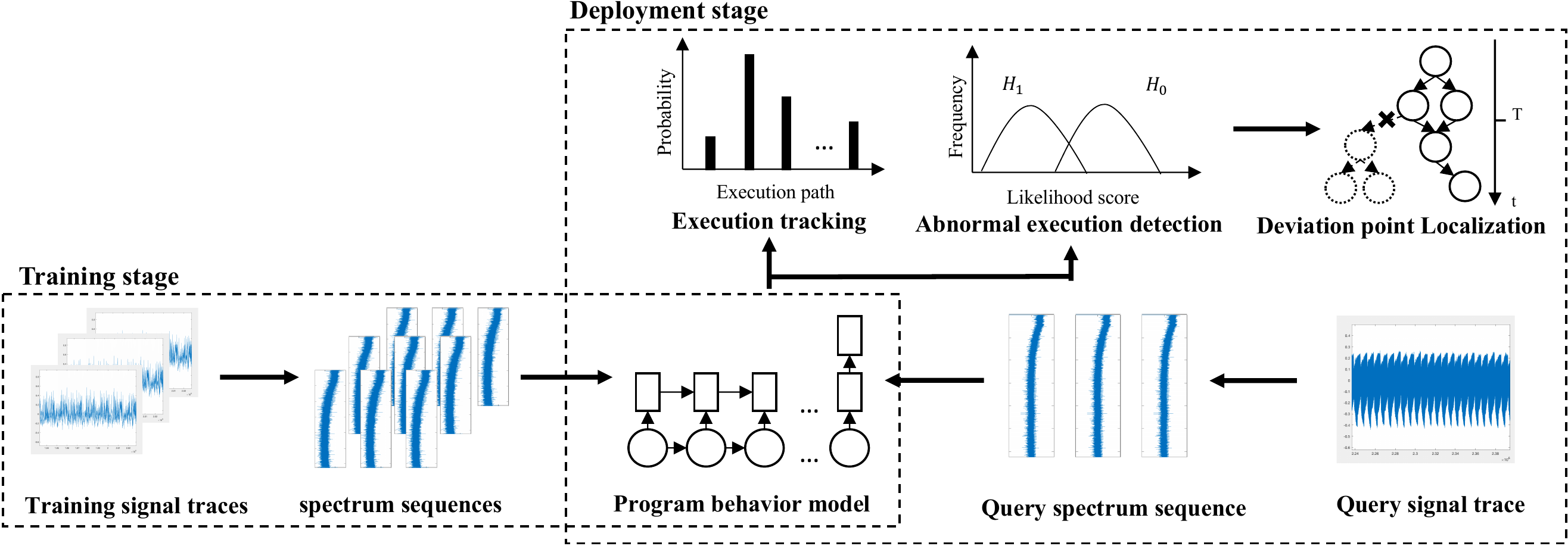}
  \caption{\name's control flow integrity monitoring.}
  \label{fig:wkflw-new}
\end{figure*}

During the PLC code execution, the processor clock frequency and switching of the underlying CMOS devices along with the power regulation board result in change of electric current in the PLC circuitry. The current produces time-varying magnetic field that interacts with the electric field leading to an electromagnetic (EM) wave. The EM wave propagates perpendicular to electric and magnetic fields~\cite{Agrawal2003}.  In order to radiate this EM emanation, an antenna is required. The components on the PLC's printed circuit board (PCB) act as antennas. The transmission range of these waves increases with the increase in the surface area of the antenna. These emanations from the PLC boards can be captured by an external electromagnetic sensor placed near the emanation source. 

\name uses these near-field EM waves as the PLC side channel, because they leak information about the program running on the device~\cite{genkin2016physical}. Different instructions usually expose different emanation patterns. Thus, the collected electromagnetic signal traces during program executions have unique local characteristics depending on the runtime control flows. This observation is utilized by \name to fingerprint the side-channels of legitimate program executions and identify unknown (malicious) code injections and/or control flow hijacking attempts. 

Recent attempts have been made to monitor micro-controllers such as STC89C52~\cite{liu2016code} and PIC16F687~\cite{eisenbarth2010building} based on power signal analysis~\cite{liu2016code, eisenbarth2010building} that require physical manipulation of the circuits for sensor placement. The data acquisition draws current from the controller boards potentially affecting its functionalities that triggers a red flag for practical deployment in controllers for mission-critical real-time operations. In comparison, \name's contactless, passive and non-intrusive monitoring of commercial PLC ARM processors using an inexpensive EM sensor for control flow integrity is a more challenging endeavor. 

Due to the PLC architecture, the execution times of individual instructions are not fixed to the processor's clock cycle. A list of estimated completion times for instructions is provided in the user manual. However, in practice based on our observations, even the execution time of a single instruction varies across different execution runs of the same PLC code. This makes the signal analysis using time-based truncation infeasible. Being contactless, \name has to deal with a large amount of signal noise. Our collected electromagnetic signal traces have very low signal to noise ratio (SNR) such that repeatable local patterns along the execution paths (leveraged by \cite{liu2016code}) cannot be observed in the time domain. 

To deal with these problems, \name borrows ideas from speech recognition research~\cite{rabiner1989tutorial}. \name looks at frequency representations of signal sections within a local sliding window. \name extracts signal segments via a sliding window on the collected signal. Each segment consist of several consecutive instructions. \name then computes the power spectral density of each segment.

Unlike time domain signals, we observed that the patterns in the frequency representations are much more stable and robust to noise. This is because the local spectra (spectrum sequence) are computed through weighted summation of all time signal points within the window. Hence, the white noise is not cumulated, while the underlying desired deterministic signal is. Therefore, a sequence of local spectra extracted from the PLC code execution EM signal trace includes repeatable patterns to characterize individual execution paths. \name deploys the aforementioned analysis to model the execution behavior of target PLC programs.

For completeness of the results, signal traces of all feasible control flows of the program are collected. \name monitors the network link between the HMI servers and the PLC controllers, and intercepts the control logic uploads to the PLC. Through symbolic execution~\cite{mclaughlin2014tsv}, the execution path predicates are aggregated and checked by an SMT solver for satisfiability. Consequently, infeasible executable paths are eliminated. 

For each remaining feasible path, \name calculates several concrete test cases through counterexample-guided inductive synthesis~\cite{solar2008program}. More specifically, to calculate the first test case for a path, its aggregated path condition expressed as a conjunctive logical expression $\varphi_p = (\varphi_1 \wedge \varphi_2 \wedge \cdots \wedge \varphi_n)$) is fed to the SMT solver. The solvers produces a concrete input value set (e.g., $i = 20$). Calculating the second concrete input for the same path involves feeding $\varphi_p \wedge \neg\varphi_i$ to the SMT solver, where $\varphi_i := (I == 20)$ and $\neg$ represents logical negation. The next concrete inputs are calculated similarly.

\name runs the program on the PLC using the generated test cases for each execution path, and collects the electromagnetic emanations using an external sensor. The collected signal traces along with their labels (corresponding control flows) are fed to a sequence neural network classifier for training. All these steps are performed offline. During the runtime, \name's external sensor collects the PLC's emanations and employs the classifier to determine whether the signal trace belongs to the feasible legitimate execution paths. A modified execution path (e.g., a maliciously injected PLC program) will lead to a change in electromagnetic emanations away from the samples observed by the classifier during the training phase. The deviation triggers a red flag by the classifier, and the execution is marked as malicious. 

\section{EM-Based Control Flow Monitoring} 
\label{sec:control-flow}

There have been works utilizing electromagnetic side channel signals to detect abnormal executions~\cite{stone2013radio}. They follow a template matching scheme, where the query signal is compared with all constructed templates of the execution paths. Based on our experiments, such time domain-based signal matching techniques cannot distinguish fine-grained characteristics of complex program control flows accurately. To address this, \name constructs a sequential neural network classification model of pre-processed frequency domain data samples. The model therefore learns from control flow transitions from the training frequency data and encodes them in its network weights. This model describes the behavior of the program. 

\name maps the control flow transitions of any new legitimate data sequence with the learned transitions, modeled by the neural network, and determines a specific control flow that the observations correspond to. A data sequence with abnormal components such as unseen segments (due to code injection attacks) or invalid transitions between segments (due to control flow hijacking) will cause mismatches. Such mismatches accumulate along the sequence, cause the neural network states and thus the output of the model to deviate from expected values. 

\autoref{fig:wkflw-new} shows \name's work flow. During the offline training stage, each collected signal trace is transformed into a spectrum sequence (\autoref{sec:emanation-analysis}). The sequence neural network model is trained using spectrum sequences of all classes (legitimate control flows). After deployment, \name feeds the online spectrum sequences of collected query signal traces into the trained model. This results in a probability distribution computed by the model over all the classes. The class with the highest score is compared to a predefined threshold. If the score exceeds the threshold, \name assigns it to the query signal trace as the execution path that the program is taking currently. If the threshold check fails, it indicates a mismatch between the query signal trace and the trained model. Consequently, \name triggers an alert about an illegitimate control flow. \name provides more fine-grained reports about the mismatch regarding the execution point that the real-time control flow deviated from the legitimate expected flows. This information can be used later for detailed vulnerability discovery, e.g., how the control flow was hijacked. We consider the vulnerability analysis phase outside the scope of this paper.

\subsection{Offline Model Construction and Training}

To construct the classification model with sequential inputs, we use a long term short memory (LSTM) network layer~\cite{gers2000learning}. LSTM is a variation of the recurrent neural network (RNN) used for modeling sequential data. LSTM takes a sequence of inputs and maintains a hidden state vector along the sequence. This fits \name's use-case, where the observables are EM signals, while the hidden states represent the underlying unobserved code segments and basic blocks. 

At each time step of the input sequence, current hidden state vector is computed based on both the previous hidden state vector and the current input. The hidden state carries long term dependency between time steps. This enables \name to capture contextual information in the sequential control flow transitions. 

Let $[x_1, x_2, ..., x_N]$ be a spectrum sequence computed for a collected EM signal trace. $x_t$ indicates the input in the sequence at time $t$. The vector size $N$ depends on the execution time of the control flow. The current hidden state $h_t$ can be computed as:

\begin{equation}
h_t = o_t * \tanh(c_t),
\end{equation}
where $\tanh$ is the hyperbolic tangent activation function; $*$ denotes the entry-wise product; $o_t$ is the output gate vector; and $c_t$ is the cell state vector. The two vectors can be computed as:

\begin{equation}
\label{eq:vectors}
\begin{split}
o_t &= \text{sigmoid}(W_ox_t+U_oh_{t-1}+b_o) \\
c_t &= f_t * c_{t-1} + i_t * \tanh(W_cx_t+U_ch_{t-1}+b_c).
\end{split}
\end{equation}

In \autoref{eq:vectors}, $W_o, U_o, b_o, W_c, U_c, b_c$ are the weights of the neural network units. Note that the design extends the basic network architecture described in \autoref{sec:background}. $c_{t-1}$ and $h_{t-1}$ are state vectors passed from the previous time step. $f_t$ and $i_t$ represent the forget and input gate vectors, respectively. These vectors are designed to keep only useful contextual information and acquire new information: 

\begin{equation}
\begin{split}
f_t &= \text{sigmoid}(W_fx_t+U_fh_{t-1}+b_f) \\
i_t &= \text{sigmoid}(W_ix_t+U_ix_t+b_i),
\end{split}
\end{equation}
where $W_f, U_f, b_f, W_i, U_i, b_i$ are the unit weights. We add a dense layer followed by a softmax function after the output of the LSTM layer at the last time step $h_N$. This maps the network to a probability distribution over all legitimate control flows in the PLC code.

\begin{equation}
p = \text{softmax}(Wh_N+b),
\end{equation}
where the height the weight matrix $W$ is the same as number of legitimate control flows.

Intuitively, the neural network model output is a one-hot~\cite{uriarte2014one} vector $q$. It has a $1$ on its entry that corresponds to the identified control flow and $0$s on all its other entries. $q$ can also be viewed as a probability distribution. We define the loss function of our model as the cross entropy between the model's actual output $p$ and the target vector $q$. The resulting cross entropy measures the difference between two probability distributions:

\begin{equation}
l = -\sum_{i} p_i \log q_i,
\end{equation}
where $i$ is the index of the legitimate control flows. The total loss of our model is the sum of losses over all the training EM spectrum sequences and their corresponding class labels (control flows). During the training, weights of the model are tuned iteratively as described in \autoref{sec:threat-model}. A well-trained model will have its output probability distribution very close to its corresponding one-hot vector, and the output distribution will bias remarkably towards the corresponding control flow. 

\begin{figure}[tp]
  \centering
  \includegraphics[scale=0.3]{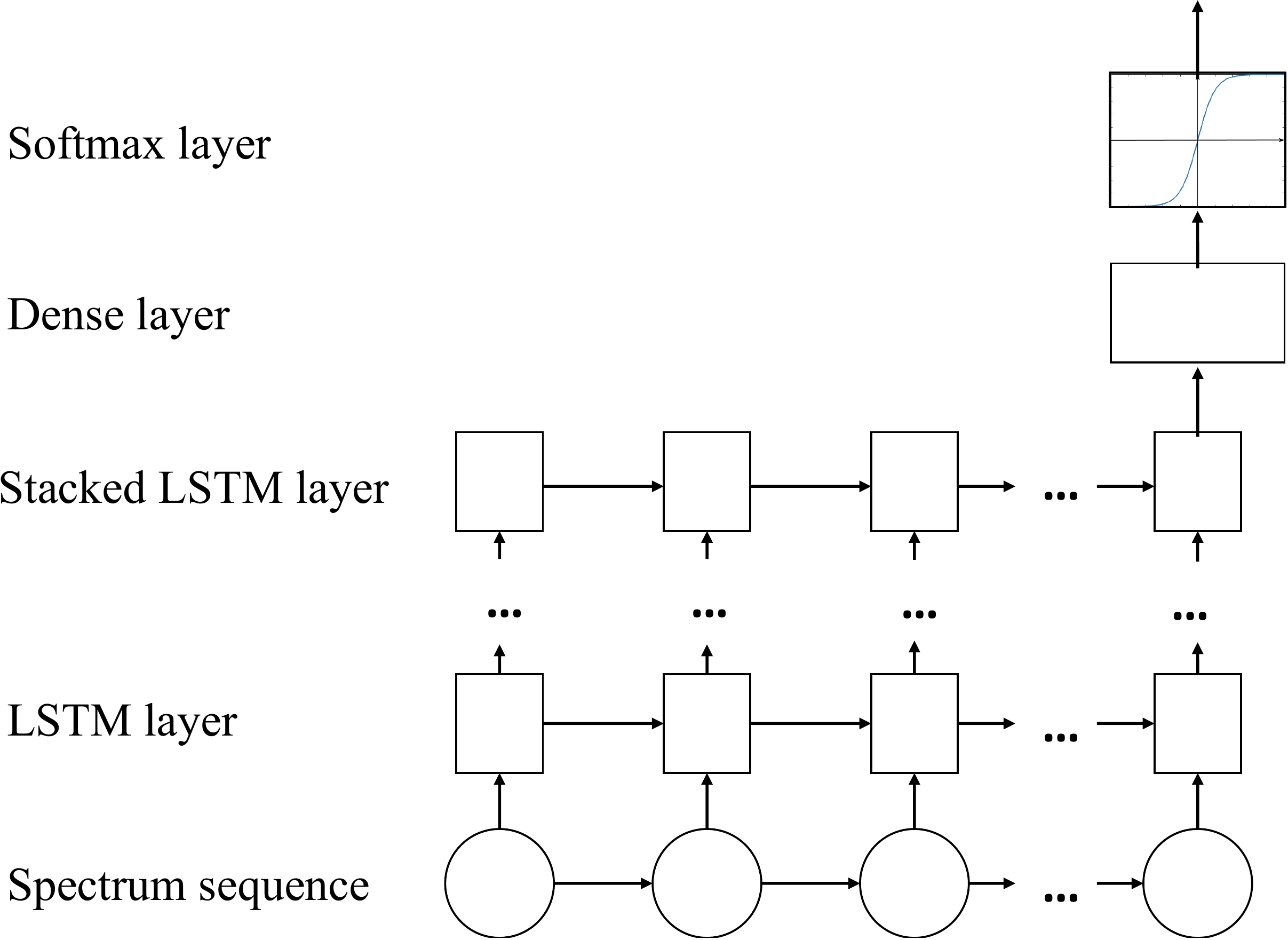}
  \caption{Network architecture of proposed model.}
   \label{fig:arch}
\end{figure}

The overall architecture of \name's network model is shown in \autoref{fig:arch}. The size of the hidden state vector $h_t$ can be increased to carry more information along the sequence. Moreover, multiple LSTM can be stacked to be more capable of characterizing the EM spectrum sequences for PLC execution classification. However both adjustments increase the computational complexity. To ensure efficient online classification, we kept the network model size minimal as long as it did not affect \name's accuracy of malicious execution detection.

The collected electromagnetic signals suffer from random perturbations caused by EM interference of other components on the PLC device. To reduce the noise effect, \name provides the neural network model training with many EM signal traces for each control flow. This is possible through \name's PLC code analysis and generation of several test-cases for each feasible execution path. Consequently, the neural network training algorithm receive many traces with the same label (control flow) each incorporating random noise. This enables the neural network's data-driven feature extractions procedures to train its unit weights based on the main signal ignoring the noise margins. 

\subsection{Online PLC Execution Monitoring}
\label{sec:online-monitoring}

\name uses the trained model at runtime to detect anomalous executions. EM spectrum sequences from legitimate executions will have their network outputs distribution heavily biased towards the corresponding class label. The class with the highest probability will be reported as the identified control flow $I$:
\begin{equation}
I = \arg\max_i p_i,
\end{equation}
and its likelihood score can be calculated as $L = \max_i p_i$. $L$ represents how likely this signal trace belongs to the legitimate PLC code. In the case of correct classification outcome, the network's input transitions match with the corresponding control flow transitions of the PLC control logic and the desired network state is maintained as the most likely state along the input EM trace. 

Malicious control flows constitute either execution of a maliciously injected new code or code reuse attacks that execute the available instructions while deviating from legitimate control flows at some point. The introduced new instructions or the control flow deviation cause a mismatch between the observed EM signals and the neural network's learned transitions. This reduces the bias in the neural network model's calculated probability distribution increasing its entropy. Therefore, by setting a threshold on the likelihood score $L$, abnormal executions can be identified as they match none of the known legitimate control flows. 

Let $H_0(H_1)$ indicate the legitimate (malicious) execution, the detection problem can be expressed as:

\begin{equation}
L\stackrel[H_0]{H_1}{\lessgtr}\varepsilon,
\label{eq:detection}
\end{equation}
where $\varepsilon$ is the preset threshold.

For malicious executions, \name can also locate the point, where PLC execution deviated from the legitimate flows. Let $h = [h_{n1}, h_{n2}, \cdots, h_{nN}]$ be the hidden state sequence of the $n$-th LSTM layer of our model for a query EM spectrum sequence input for a malicious execution. Let's also assume \name identifies the hidden state sequence $h^I = [h^I_{n1}, h^I_{n2}, \cdots, h^I_{nM}]$ as the most likely legitimate control flow that corresponds to a given query EM trace. The deviation point can be located by computing the distance $d_t$ between the two sequences at each time step $t$:

\begin{equation}
d_t = \sqrt[]{(h_{nt}-h^I_{nt})^2}.
\end{equation}

The deviation of the PLC execution from the legitimate control flows is reflected in the sudden change of signal traces and thus the neural network inputs. A changed spectrum input will cause its corresponding hidden state vector to move away from its expected vector in the state space. Therefore, a sudden step increase in the distance sequence $[d_1, d_2, \cdots]$ indicates the point, where the deviation happens.

\section{Implementation and Evaluation} 
\label{sec:evaluation}

We evaluated \name on real-world settings with commercial PLC devices and using legitimate and malicious control logics. We first describe our experimental setup including the signal acquisition system and the target PLC model. We will discuss the results on the electromagnetic emanations of the target PLC and their discriminative spectrum characteristics. We measure \name's accuracy in classifying legitimate control flows, and detecting malicious executions. We compare \name's data-driven approach with traditional model-based solutions using hidden Markov models~\cite{liu2016code} We finally test the performance of \name on several real applications. An Intel i7-6800K CPU was used to compute frequency representations, and HMM training and testing. Our LSTM neural network model was trained on an NVIDIA GTX1080 GPU.

%\input{Sources/subsec-implementation}

%%%%%%%%%%%%%%%%%%photo for experimental setup %%%%%%%%%%%%%%%%

\begin{figure}[tp]
  \centering
  \includegraphics[width=.47\textwidth]{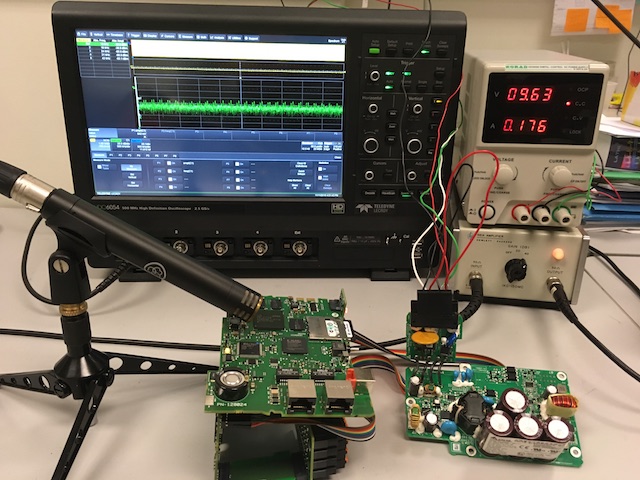}
  \caption{Experimental setup including the PLC, external sensing probe, the amplifier, and the sampling oscilloscope.}
  \label{fig:ph-exp-stup}
\end{figure}

\begin{figure}[tp]
  \centering
  \includegraphics[width=0.47\textwidth]{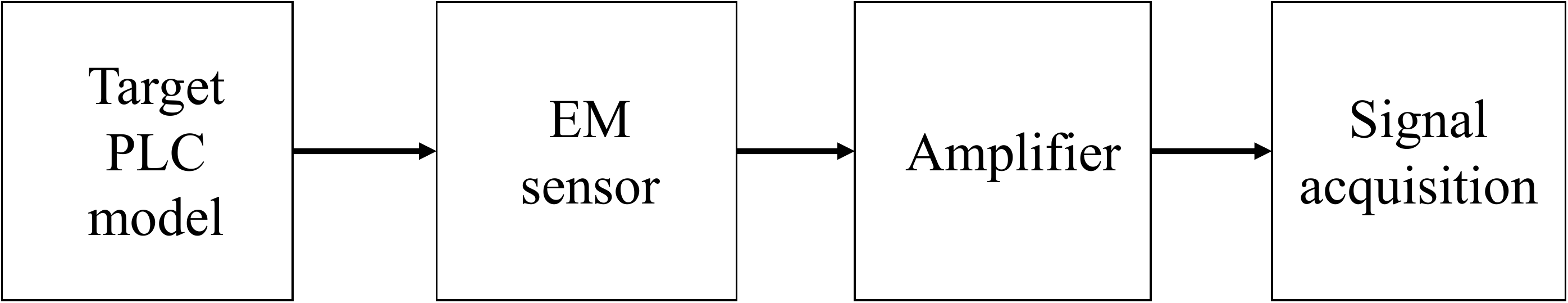}
  \caption{Experimentation test-bed configuration for electromagnetic (EM) side channel analysis.}
  \label{fig:exp-setup}
\end{figure}

\begin{figure}[tp]
  \centering
  \includegraphics[width=0.2\textwidth]{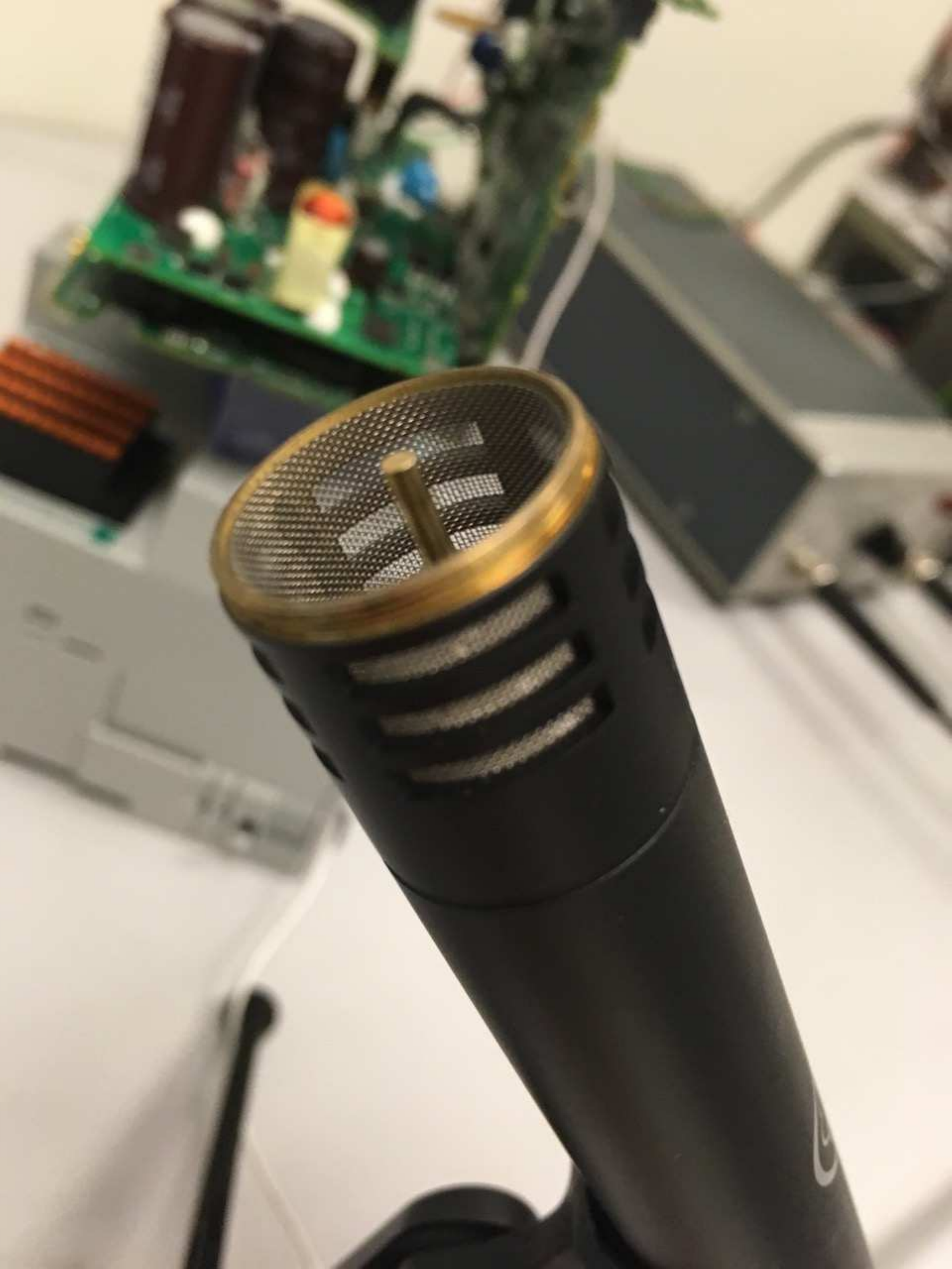}
  \caption{AKG P170 condenser microphone without transducer serving as an electromagnetic probe.}
  \label{fig:mic}
\end{figure}

\subsection{Experimental Setup}
\label{sec:setup}

\autoref{fig:ph-exp-stup} shows our signal recording setup that consists of a recording probe and an amplifier. The corresponding test-bed configuration and component interconnection is shown in  \autoref{fig:exp-setup}. In our experiments, we used the  Allen Bradley 1769-L18ER-BB1B CompactLogix PLC (with ARM Cortex-M3 processor). Allen Bradley PLCs are the most popular and widely used controllers in many industrial control systems in North America. We used an AKG-P170 condenser microphone without the acoustic capsule or transducer (\autoref{fig:mic}) as an antenna to receive the electromagnetic emanation\footnote{When the acoustic capsule is detached from the microphone, the remaining part of the microphone serves as an antenna since the coil in the microphone is sufficiently long to receive the signals emitted from the board.}. 
%%%%% passive and active antenna
We also tried a Langer LF-R 400 passive antenna. However, we achieved the best signal sensitivity from the AKG-P170 microphone. This is because the AKG-P170 microphone together with the phantom power supply can be viewed as an active antenna. Active antennas have better sensitivities than passive antennas, since signals are pre-amplified.

We used an HP-461A amplifier to increase the signal strength. %sampled the signal at different frequencies from \textcolor{green}{10 $MHz$} to \textcolor{green}{100 $MHz$}, and 
We observed that most of the informative frequency variations appear below 5 $MHz$. Accordingly, we set our sampling rate to preserve most frequency information while maintaining a moderate computation complexity for online malicious code detection performance.

\subsection{PLC Electromagnetic Emanations}

We performed numerous tests and inspected various regions of the three PLC PCB boards to identify the point that emits the most distinguishable EM signals. Once that point was identified, we adjusted our directional EM probe to focus on the point while collecting the EM emanations for our  experiments. %

\begin{figure}[ht]
\centering
\begin{subfigure}[b] {0.47\textwidth}
  \includegraphics[width=\textwidth]{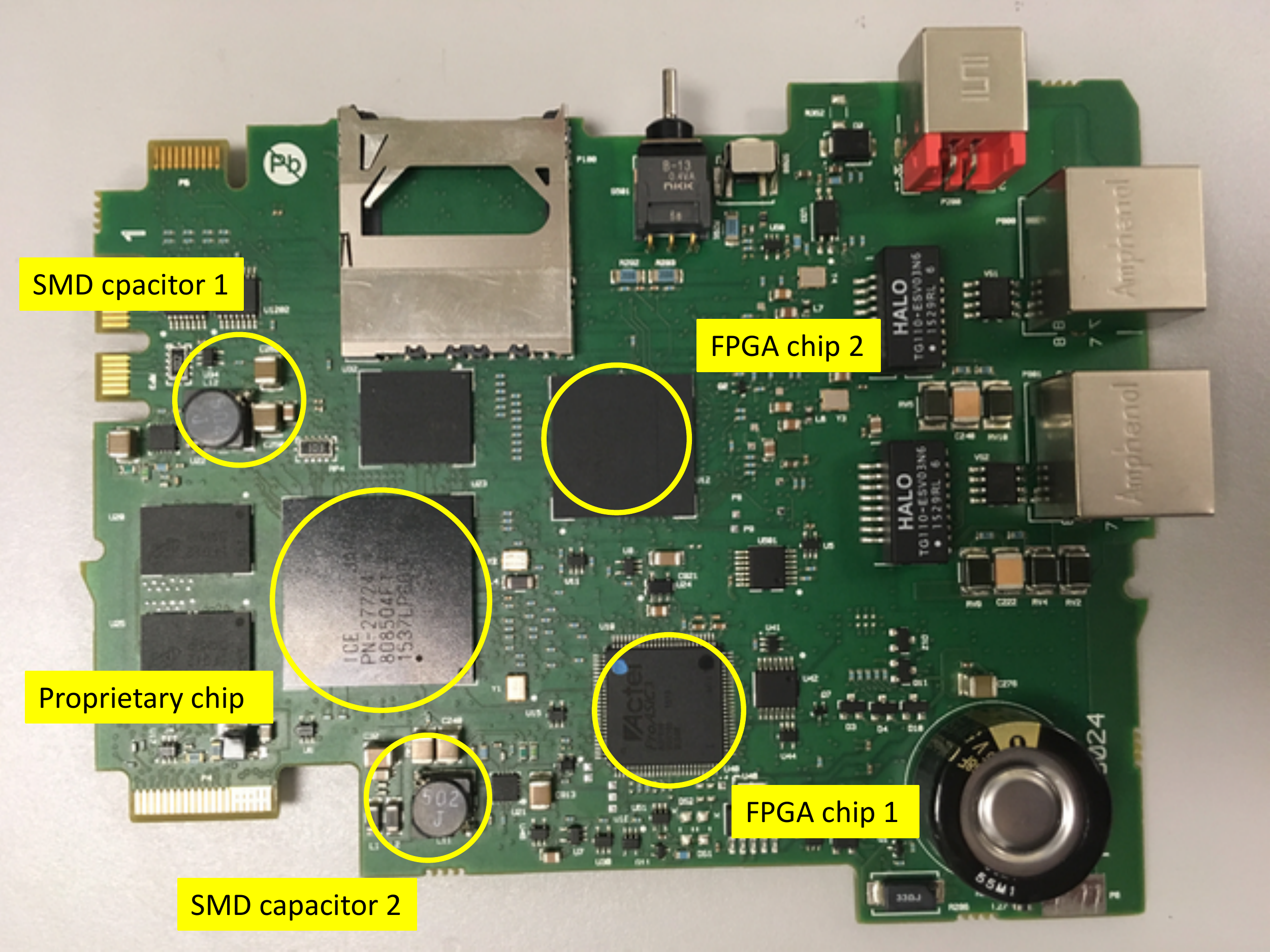}
  \caption{One of the PLC's three PCB boards: surface area of the propitiatory chip is larger compared to other chips.}
  \label{fig:ph-locations}
\end{subfigure}

\vspace{0.15in}

\begin{subfigure}[b]{0.47\textwidth}
  \includegraphics[width=\textwidth]{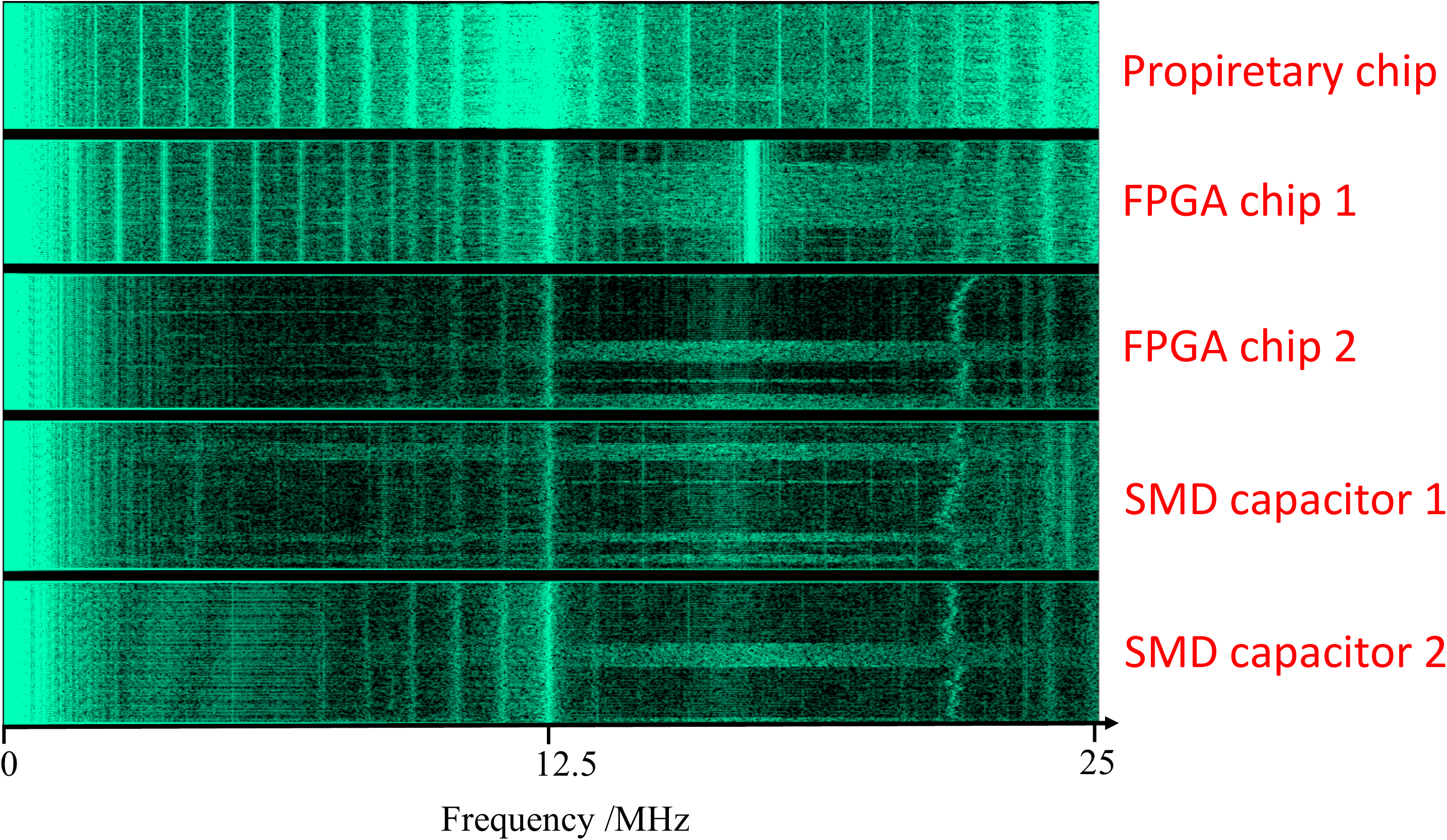}
  \caption{Spectrogram for different locations.}
  \label{fig:locations}
\end{subfigure}

\vspace{0.15in}

\begin{subfigure}[b]{0.47\textwidth}
  \includegraphics[width=\textwidth]{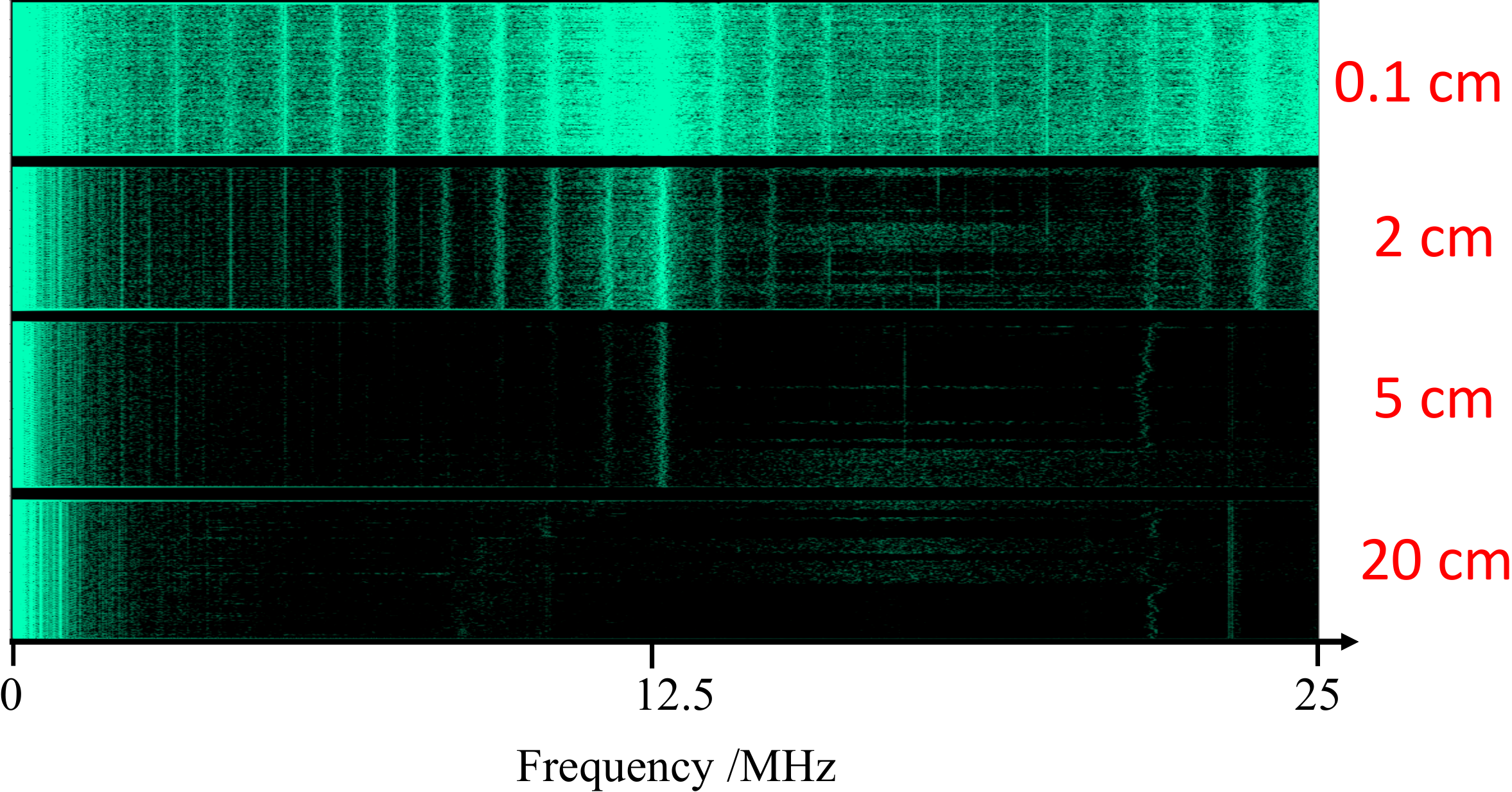}
  \caption{Spectrogram for different distances.}
  \label{fig:distance}
\end{subfigure}
\caption{EM emanation by the PLC's communication board.}
\label{fig:board}
\end{figure}

%Different instructions result in  different time varying currents in the circuit. The currents generate electromagnetic emanations. Any conductor on the PLC board can act as a transmit antenna; the larger the conductor area the stronger  the transmitted signal .  %
%
%
\autoref{fig:ph-locations} shows the components that we mainly investigated. The main sources of emanation were the proprietary Allen Bradley chip, the field-programmable gate array (FPGA) and the surface mount device (SMD) capacitors on PLC's communication PCB board. The SMD capacitors are involved in the voltage regulation for the chips. % 
\autoref{fig:locations} shows the strength of the corresponding emanation from each point. Since the surface area of the SMD capacitors is very small, the corresponding emission was rather weak. The surface area of the Allan Bradley chip is relatively larger, and hence the corresponding emission was stronger. We proceeded by focusing on that chip for our following experiments.  \autoref{fig:distance} shows how the captured signal appears with as the probe-chip distance increases.
%%% location and distance of the collected signals
The EM signals were collected by the probe located 0.1 cm away from the proprietary chip.

\begin{figure*}[ht] 
\centering
\begin{subfigure}[b] {0.46\textwidth}
  \includegraphics[width=\textwidth]{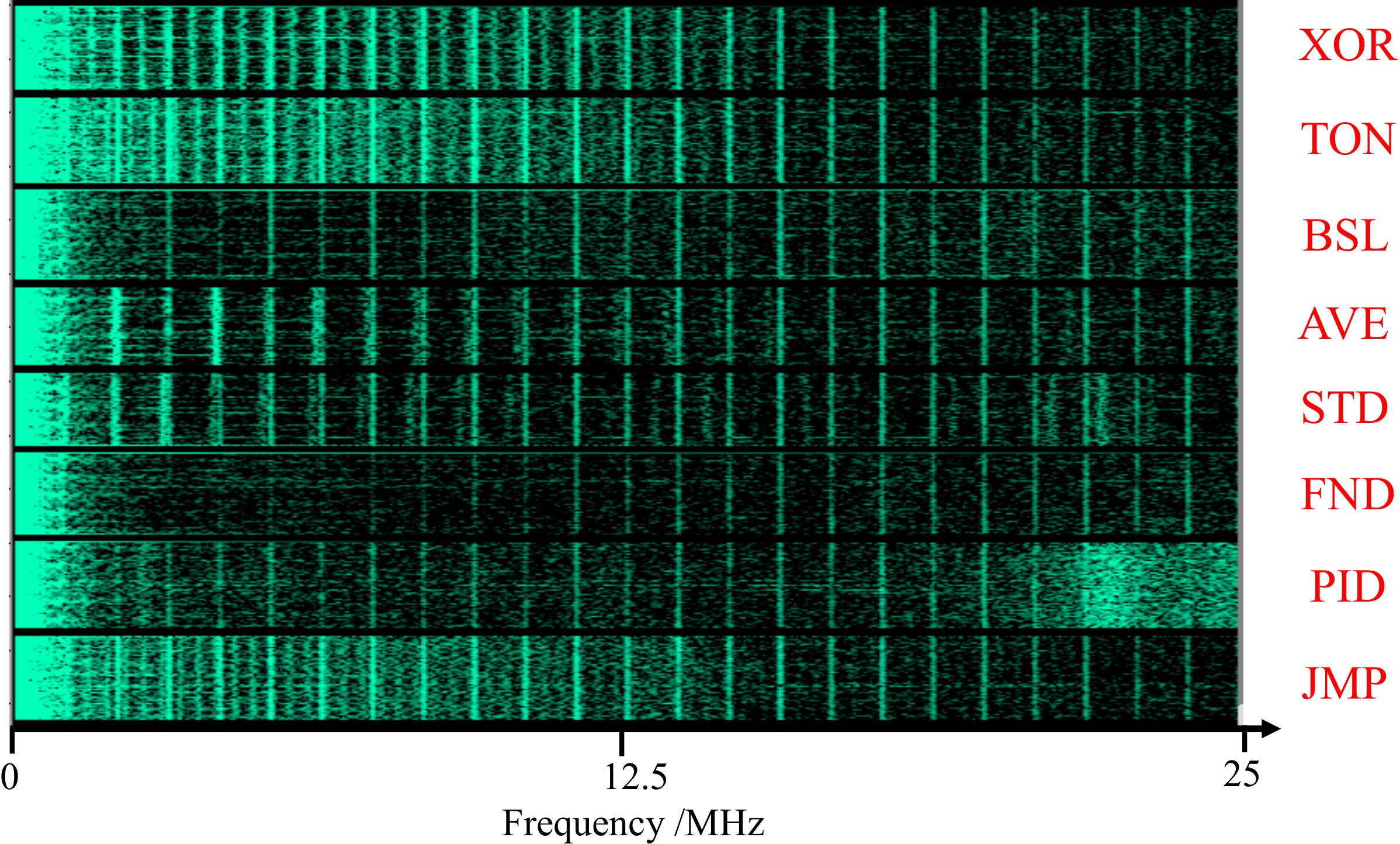}
  \caption{Pattern for PLC instructions.}
  \label{fig:16ins}
\end{subfigure}
\qquad\qquad
\begin{subfigure}[b]{0.46\textwidth}
  \includegraphics[width=\textwidth]{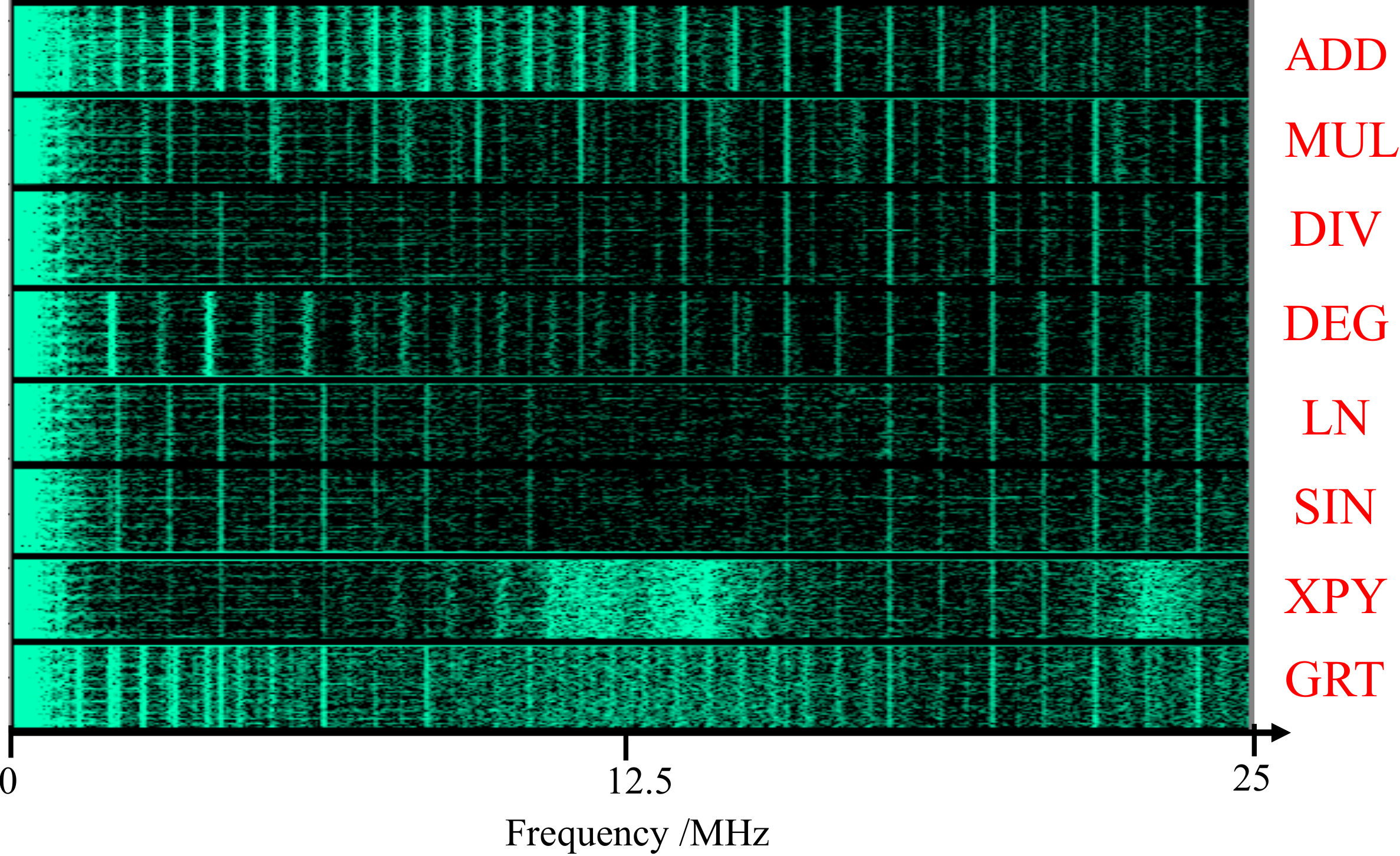}
  \caption{Pattern for PLC instructions.}
  \label{fig:16ins2}
\end{subfigure}
\caption{Spectrogram patterns of PLC instructions.}
\label{fig:spectralpatterns}
\end{figure*}

We investigated the differences among the EM emanations from the PLC execution of different instruction types.  Different PLC instructions have different execution times and computation complexities thus different power consumptions that is reflected in the emanation signals as discriminative spectral patterns. \autoref{fig:spectralpatterns} shows the results. These spectral pattern types are the core basis for \name's design.

\autoref{fig:spectralpatterns} visualizes the discriminative spectral patterns of different PLC instructions. PLC's (ARM) ISA include 22 different types of instructions. We show the results for only the 16 types that are commonly used in PLC programs\footnote{The instructions include arithmetic instructions (ADD, MUL, DIV, DEG), advanced math instructions (LN, SIN, XPY, STD), comparing instructions (XOR, GRT), array manipulation instructions (BSL, AVE, FLL) and control instructions (TON, JMP).}. Modules involving complicated computations, such as PID\footnote{This PLC programming module implements the proportional-integral-derivative (PID) control algorithm~\cite{aastrom2006advanced}.} were also tested. \autoref{fig:spectralpatterns} shows that different instructions give rise to EM signals  with different intensities at different frequencies. \name exploits these fingerprints to estimate the PLC's internal runtime execution state and dynamic control flow using the collected EM emanations. 

\begin{table*}[tp]
\centering
% \footnotesize
\caption{Confusion matrix for the classification.}
\begin{tabular}{  c | c  c  c  c  c c } % center
	 & ADD & SIN & XOR & BSL & JMP & PID\\ % LINE1
	\hline
  ADD & \textbf{78.63\% (747)} &  5.26\% (50) & 8.21\% (78) & 1.58\% (15) & 4.95\% (47)  & 1.37\% (13)\\
% 	\hline
  SIN & 5.36\% (56) & \textbf{83.54\% (873)} &  5.65\% (59) &  1.05\% (11) &  2.39\% (25) & 2.01\% (21) \\
% 	\hline
 XOR &  8.13\% (75) & 7.48\% (69) & \textbf{69.31\% (639)} &   0.11\% (1) & 12.47\% (115) & 2.49\% (23) \\
% 	\hline
 BSL & 1.24\% (12) & 1.65\% (16) &  0.10\% (1) & \textbf{95.24\% (921)} &  0.21\% (2) & 1.55\% (15) \\
% 	\hline
 JMP & 5.44\% (53) & 3.29\% (32) & 12.32\% (120) &  0.21\% (2) & \textbf{76.49\% (745)} & 2.26\% (22)\\
% 	\hline
 PID & 0.32\% (3) & 2.11\% (20) & 2.64\% (25) &  0.21\% (2) & 1.27\% (12) & \textbf{93.46\% (886)} \\
\end{tabular}
\label{fig:classificationvalidation}
\end{table*}

% \autoref{fig: spectralpatterns} shows the PLC's EM emanations for sample instructions\footnote{The instructions include arithmetic instructions (ADD, MUL, DIV), advanced math instructions (SIN, LN), comparing instructions (XOR, GRT) and memory access instructions (BSL, AVE, FND).} of the 16 types that are commonly used in PLC programs. Modules involving complicated computations, such as PID\footnote{This PLC programming module implements the proportional-integral-derivative (PID) control algorithm~\cite{aastrom2006advanced}.} were also tested. The figure shows that different instructions give rise to EM signals  with different intensities at different frequencies. \name exploits these differences (fingerprints) to estimate the PLC's internal runtime execution state and back-to-back branch traversals using the collected EM emanations. 

%%%%%%%%%%%% location and distance

%%%%%%%%%% instruction emanations

%%%%%%%%% emanations while programming and booting
\begin{figure} [ht] 
\centering
\begin{subfigure}[b] {0.47\textwidth}
  \includegraphics[width=\textwidth]{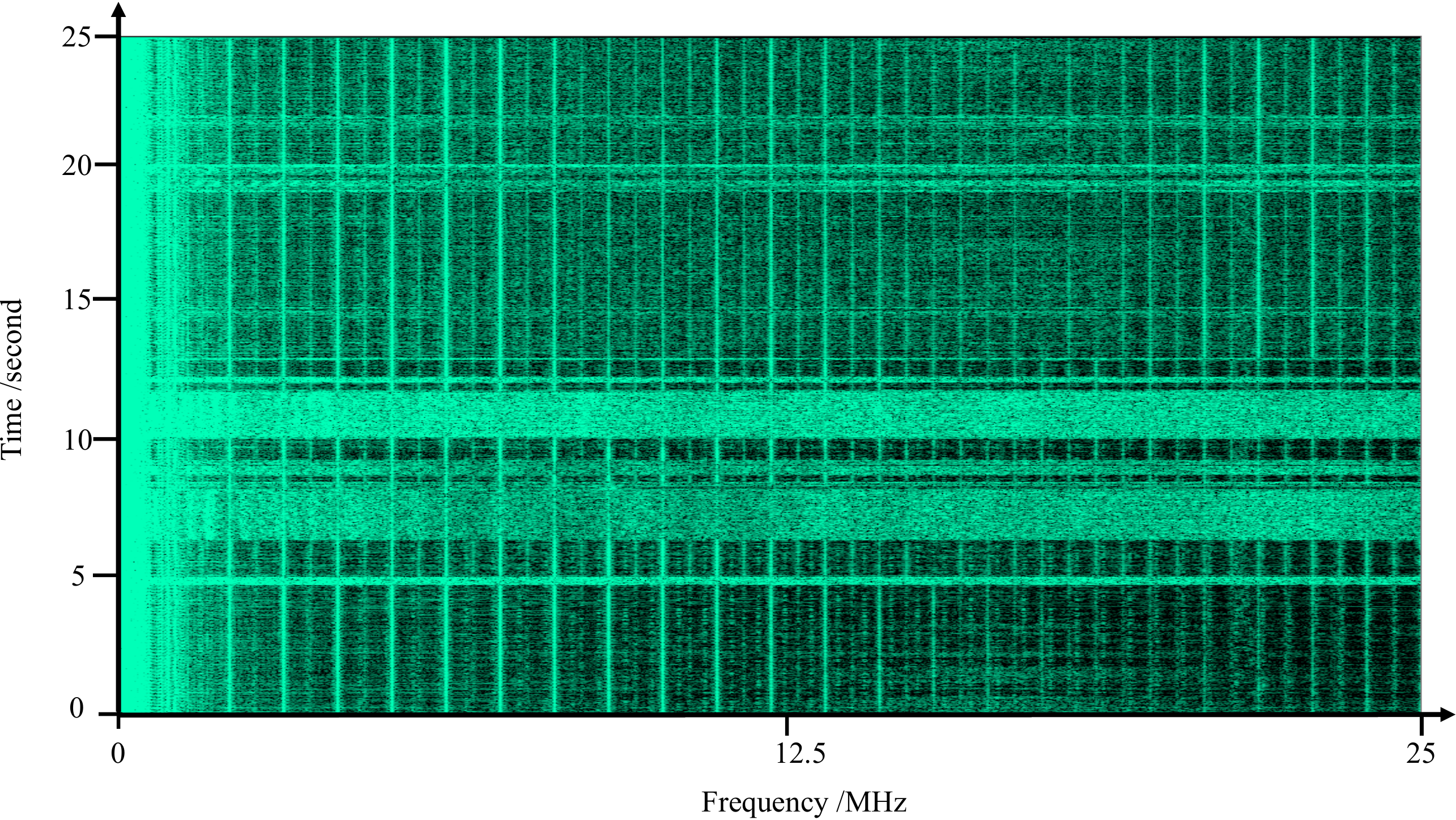}
  \caption{Spectrogram during uploading of PLC program.}
  \label{fig:uploading}
\end{subfigure}
\begin{subfigure}[b]{0.47\textwidth}
	\includegraphics[width =\textwidth]{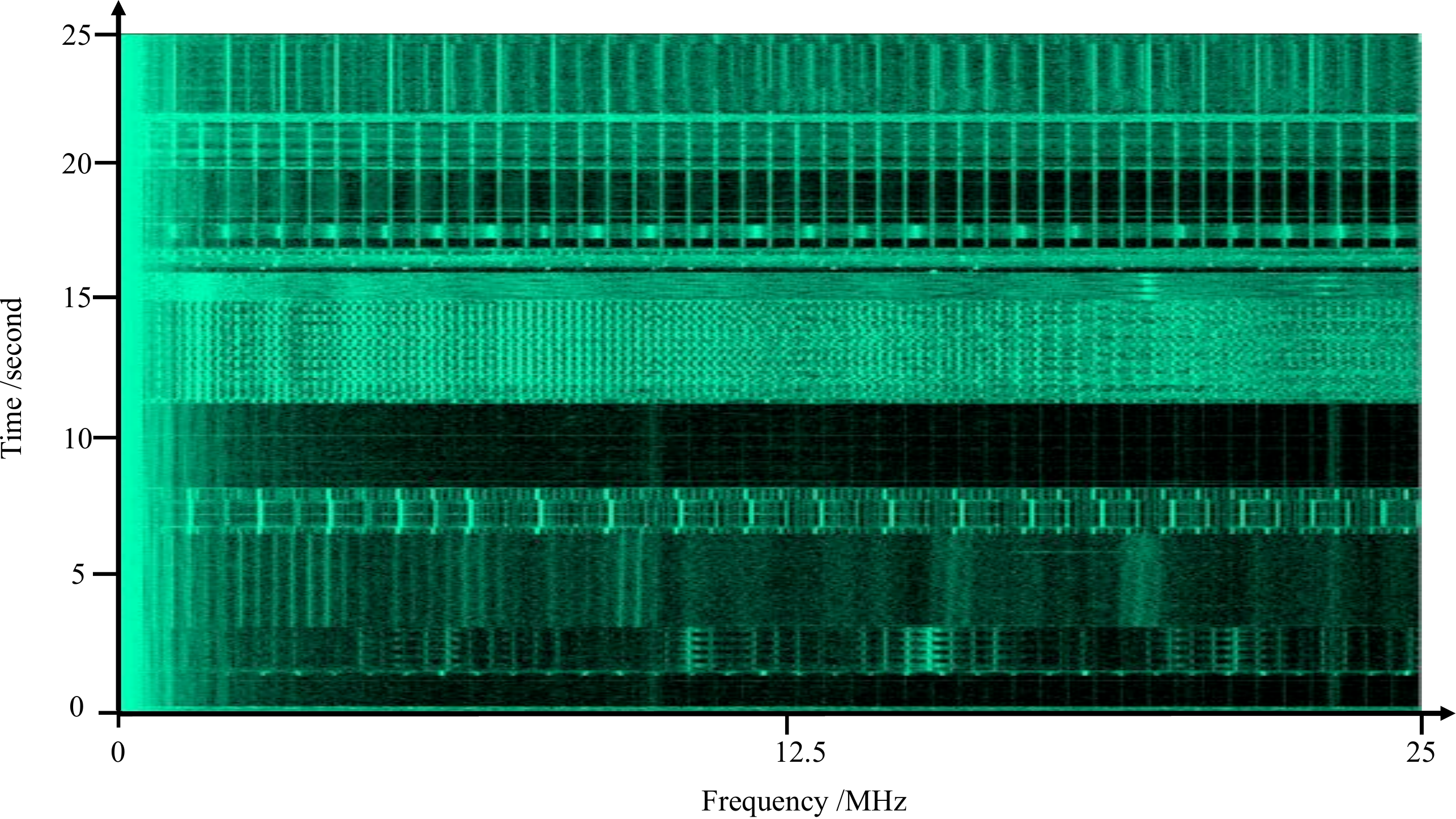}
	\caption{Spectrogram during the booting phase of the PLC.}
	\label{fig:booting}
\end{subfigure}
\caption{Spectral patterns of PLC instructions.}
\end{figure}

We further verify our visual observations by performing a classification validation on spectra of emanation signals of these instructions using the random decision forests algorithm~\cite{ho1995random}. We used Weka~\cite{hall2009weka} to implement the classifier. One instruction of each instruction type (\autoref{fig:spectralpatterns}) was tested. An emanation signal of 200 $\mu s$ was collected under sampling rate of 50 $MHz$ and transformed into spectrum representation. 1000 such signal traces were collected for each instruction type to train the classifier, and the same number of traces were collected for validation testing. 

\autoref{fig:classificationvalidation} shows the classification confusion matrix. Most signals are correctly classified correctly as their actual instruction type. This shows that the spectral patterns of different types of instructions are indeed discriminative and can be used for control flow integrity monitoring. 

\name applies a sliding window to the collected emanation signals. Because of various instruction execution times, sliding windows of the same length at different points of the signal cover different combinations and number of instructions. This helps for spectral patterns of the signal segments within different windows across the signal trace to be distinguishable. Therefore, the spectra of these local signal segments characterize the emanation signals, the execution paths. \name utilizes this to construct the program behavior model.

\autoref{fig:uploading} shows the EM emanation (between seconds 6 and 12) while a control logic program is installed for execution on the PLC. The visible EM emanation pattern can be used to detect runtime (malicious) control logic uploads similar to Stuxnet~\cite{stuxnet-2010}. %We performed similar experiments and other code/data read/write operations from/to the PLC memory can be distinguished and are detectable.
\autoref{fig:booting} shows the electromagnetic emanation patterns during the PLC's boot-up process. These patterns can be used to detect when  the PLC is remotely rebooted by an adversary.

\begin{table*}[tp]
	\small
    \caption{Evaluation programs and descriptions.}
	\begin{tabular}{ | l | l | l | l | c | }
    	\hline
		 \textbf{Class} & \textbf{Name}  & \textbf{Description} & \textbf{Example applications} & \textbf{Average length (msec)} \\
        \hline
        Vector arithmetic & Matrix  & Matrix multiplication & Sensor array data processing  & 3.3 \\
        \cline{2-5}
         & Q-sort  & Quick sort & Value searching, element uniqueness &   2.1 \\
        \hline
        Numerical methods & GD & Gradient descent & Power flow optimization &  5 \\
        \cline{2-5}
		 & Newton  & Newton's method  & Vehicle trajectory estimation & 3.5 \\ 
        \hline
        Signal processing & Conv  & Convolution & Signal filtering &  9.2   \\
        \cline{2-5}
         & DCT  & Discrete cosine transform & Audio lossy compression & 17.3   \\
        \hline
      	Communications  & Dijkstra  & Dijkstra's algorithm & Routing optimization in smart grid  & 11.3  \\
        \hline
        Cryptography & AES  & AES-128 encryption & Data protection, access control  &  18.1  \\
        \hline
        Control systems & PID & PID control  & Vehicle cruise control &  6.5  \\
        \cline{2-5}
         & Patfilt  & particle filter & Object Tracking, localization estimation  & 2.5  \\
        \hline
	\end{tabular}
    \label{tab:program}
\end{table*}

\subsection{Accuracy}

We evaluated \name for PLC execution monitoring, control flow classification of ten real applications, and detection of malicious code executions.
%\footnote{\textcolor{blue}{We consider the whole program replacement with a new different malicious code as the attack vector}}.  %Sampling rate was set to be $50 MHz$. 
We computed spectrum sequences and estimated the power spectral density of signal segments. The segments were extracted using sliding windows of size $200 \mu s$, with $90\%$ overlap between successive windows. 

A stacked two-layer LSTM network with 100 nodes on both layers was employed to fingerprint the execution behavior of each program. We trained the network using stochastic gradient descent (SGD). An average of 50 epochs (iterations) were required for the network to converge on the tested programs. We obtained 100 traces for every feasible control flow of each program for training the model. For each program, a 2-fold cross validation was performed 10 times to stabilize the result. 

We chose ten real PLC programs from different application domains for evaluation purposes. \autoref{tab:program} lists the control logics along with their functionalities. These programs fall in the classes of vector arithmetic, numerical methods, control algorithms, cryptography, signal processing and communications.

\paragraph{\textbf{Comparison with HMM solutions.}} We compared our LSTM network model with a traditional hidden Markov model (HMM) based program behavior modeling approach~\cite{liu2016code}. For the HMM, the observations were defined as the signal segment or its frequency representation. HMM state was defined as unique samples in the observation set. The number of HMM states was defined as a adjustable parameter. We set the HMM number of symbols to be $100$. We fit the observations of each state with a multivariate Gaussian distribution. The parameter set (HMM's transition probabilities, observation models and initial probabilities) was estimated using Baum-Welch algorithm~\cite{moon1996expectation}. We used the forward algorithm~\citep{rabiner1989tutorial} to calculate the observation sequence likelihoods. We will present the accuracy results for both \name and HMM solutions below.

We evaluated \name accuracy from two aspects: \textit{i)} execution monitoring - to determine the control flow of a running legitimate PLC code, and \textit{ii)} malicious execution detection - to detect the control flows that are not a part of the legitimate program. \autoref{tab:acc} shows the execution monitoring accuracy results.  We evaluated \name (LSTM) with both pre-processed spectrum traces (Freq) and raw time domain signals (Time) and compared the results with HMM-based solutions. LSTM using the frequency representation (Freq-LSTM) achieves almost perfect results on all the evaluated programs. 

LSTM's better results in comparison with HMM-based solutions can be explained by the following two observations. First, the \name's LSTM network architecture is able to capture long-term dependency in the input sequence. This contextual information corresponds to the control flows of the program, and hence is essential in distinguishing different execution paths. HMM models, on the other hand, assume only 1st order data dependency in the sequence, and hence miss a lot of useful information. 

Second, \name's model is able to extract discriminative features from the input due to its stacked multi layer architecture. This contributes to the classification performance. For HMM, however, the input data is directly used for parameter estimation without any feature extraction. When raw signal segments are used as inputs, both LSTM and HMM are not able to achieve good performance (\autoref{tab:acc}). This is because raw time signals contain lots of noise, so the underlying signal characteristics cannot be recognized and hence learned by the two models. The frequency representation, however, reveals the signal characteristics as the noise (low frequency) stays far from the main signal (high frequency). 

%We plot the distribution of the likelihood scores $L$ of the testing EM traces (\autoref{sec:online-monitoring}). we also compute and receiver operating characteristic (ROC) curve, which is a standard evaluation metric for detection problems whose discrimination threshold are varied. 

%, Execution monitoring corresponds to telling the true execution branch of a given legitimate signal trace, this is a standard classification problem, we report classification accuracy on all branches as the performance evaluation criterion. Intrusion detection corresponds to telling if a given signal trace is positive (abnormal) or negative (normal), we evaluate this on a standard detection problem scheme. Specifically, 

\begin{figure} [tp]
\centering
\begin{subfigure}[tp]{0.47\textwidth}
  \includegraphics[width=\textwidth]{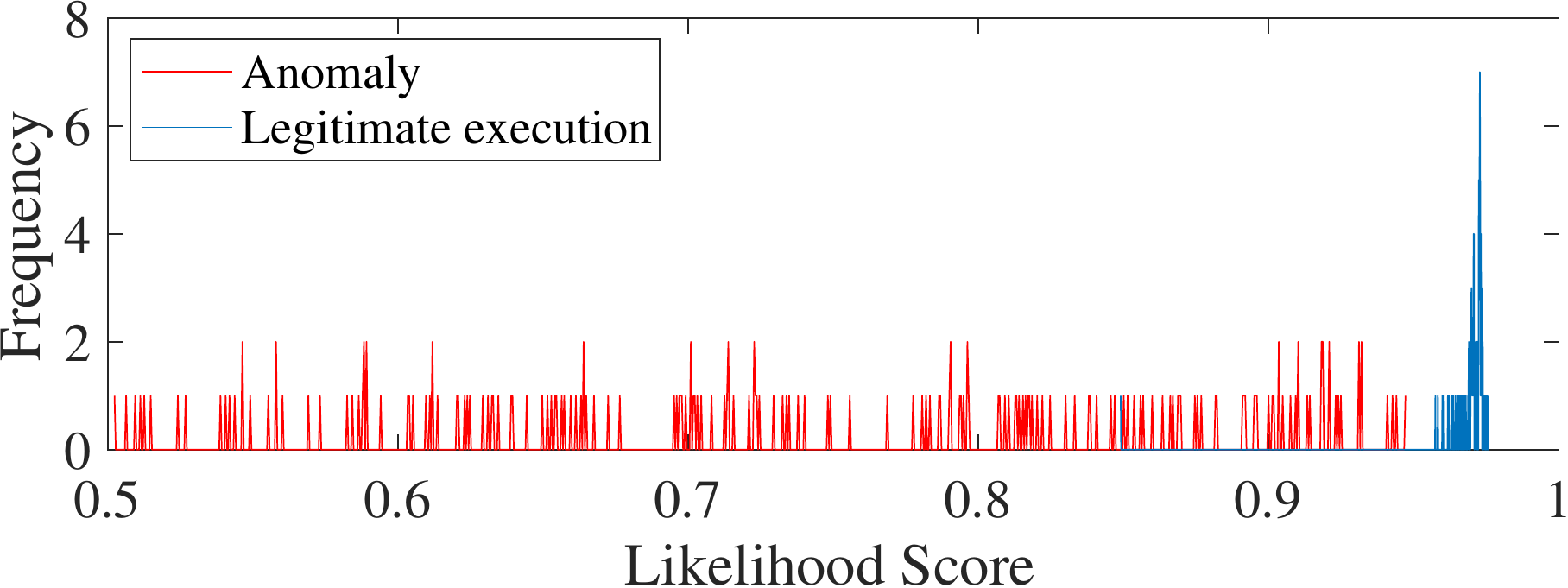}
  \caption{Newton's method}
  \label{fig:likedistnew}
\end{subfigure}

\vspace{0.15in}

\begin{subfigure}[tp]{0.47\textwidth}
  \includegraphics[width=\textwidth]{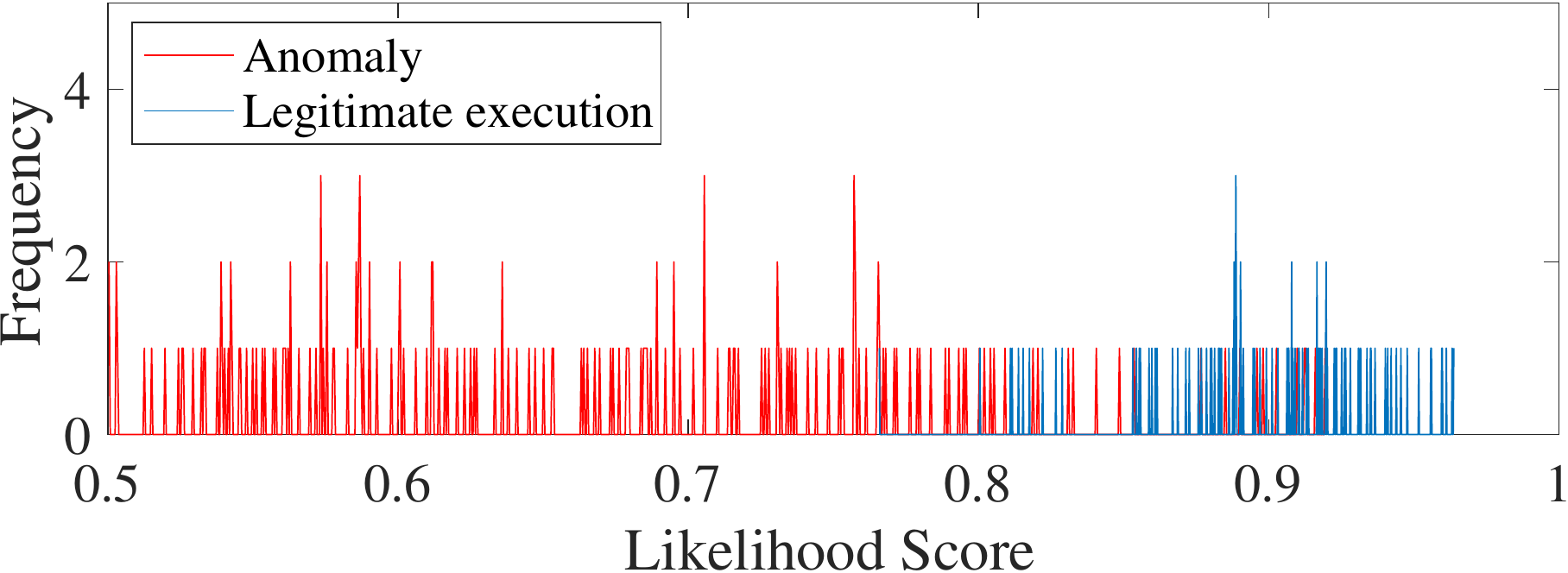}
  \caption{AES-128 algorithm}
  \label{fig:likedistaes}
\end{subfigure}
\caption{Example likelihood score distributions of the evaluated programs produced by the Freq+LSTM setting.}
\label{fig:likedist}
\end{figure}

% results for detection
For detection of malicious executions, \autoref{fig:likedist} shows the likelihood scores $L$ (\autoref{sec:online-monitoring}) of positive (abnormal) and negative (normal) samples for two example applications, namely Newton-Raphson numerical method and AES encryption algorithm. Note the negative samples tend to concentrate within a small range, while the positive samples are more spread out. This is because the number of control flows with each program is finite, and each control flow is well recognized by our network through training. Thus, the signal traces of the legitimate control flows match closely with the LSTM model. The malicious programs can take any arbitrary control flow especially in the case of malicious code injection attacks; therefore, their matching degree vary a lot.

\begin{figure}[thbp] 
\centering
    \centering
    \includegraphics[width=.47\textwidth]{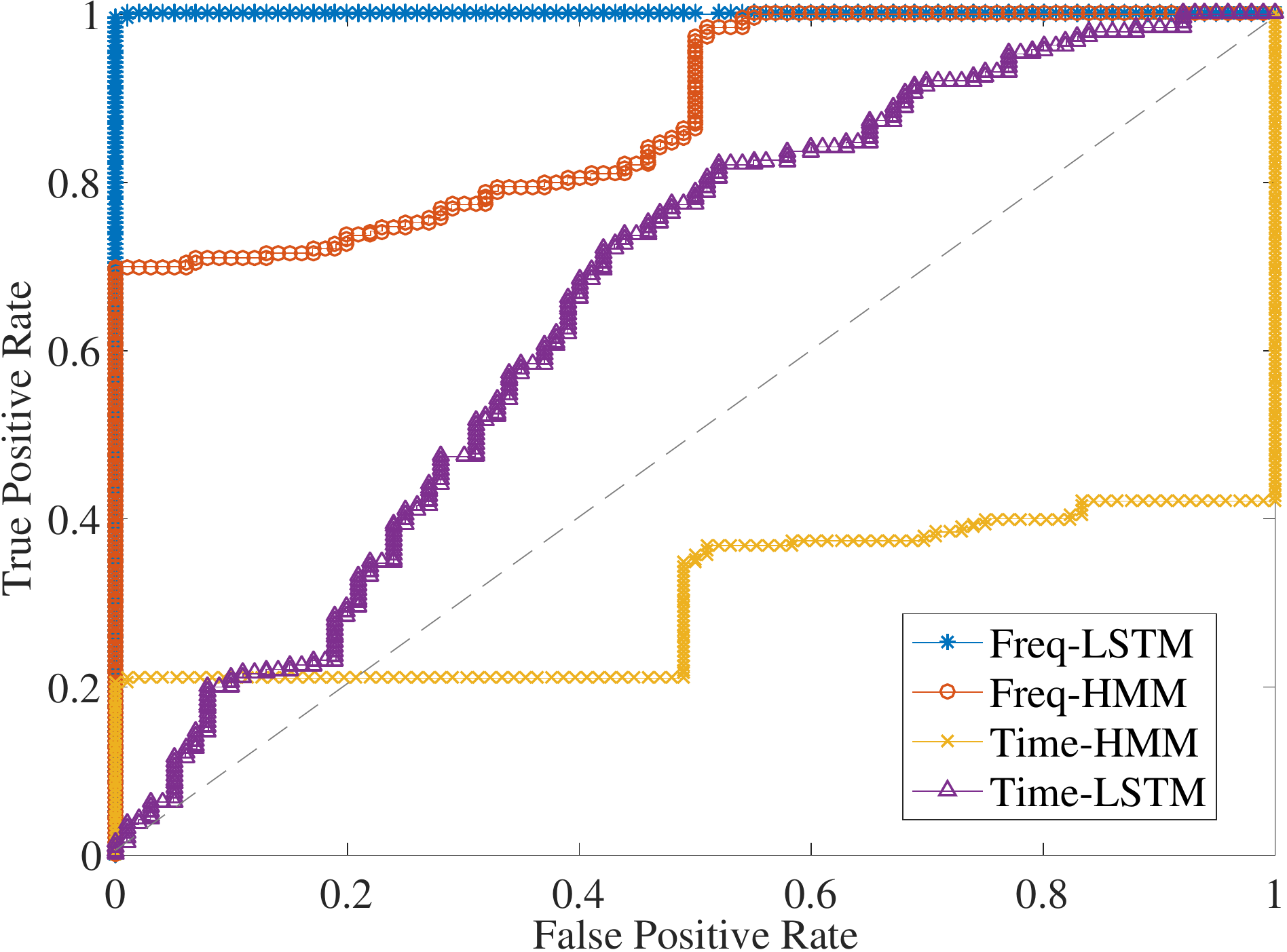}
\caption{ROC curves of all evaluated programs. AUC of the four settings: Freq-LSTM is 0.99, Freq-HMM is 0.83, Time-LSTM is 0.59, Time-HMM is 0.36.}
\label{fig:roc}
\end{figure}

\autoref{fig:roc} shows the ROC curve for the frequency and time domain data using \name's LSTM and HMM solutions. The numbers are average over all the ten applications.  \name (LSTM) using the frequency traces achieves almost perfect detection performance. Steeper ROC curve indicates better separation of positive and negative samples, and thus better performance. This is usually measured by the area under the curve (AUC). AUC is usually between 0.5 (random guess) and 1 (perfect separation). \autoref{tab:auc} shows the AUC for each target PLC program and each evaluation setup. The stacked multi-layer architecture of \name's network model captures important information both from the hidden states and the inputs, and carries it along the sequence. This results in better learning of the program behavior from the signal traces. 

Note the HMM using the raw time domain signal performs worse than random guessing (diagonal line on ROC - \autoref{fig:roc}). This is because HMM, due to its limited first order dependency assumption, is not able to characterize the temporal behavior of the signal traces well. Additionally, noisy raw time domain signal traces further contribute to its randomness and poor accuracy. 
%We look at the likelihoods of all the testing sequences for this case, the likelihoods for the negative samples are even lower than a large number of positive samples, the if we continue following the same detection criterion \autoref{eq:detection}, the results will even be worse than random guess. This is because HMM, due to its limited 1st order dependency assumption, is not able to characterize the behavior of the signal traces well, therefore the likelihoods it predicts for query signal traces don't really shows the match degree of the signal to the model, but rather a random value, thus likelihoods for legitimate executions and anomalies are mixed, therefore the detection performance is poor. Noisy raw time signal segments add to this randomness. 

We intentionally reduced the convergence threshold for the neural network's training that led to larger number of training iterations. The main reason is \name's goal to detect malicious executions and not only to classify (previously seen) legitimate control flows. The increased number of iterations resulted in more discriminatory classification outcomes, i.e., more biased probability distribution over the classes (legitimate control flows) and larger likelihood score $L$. Hence, we were able to increase the classification threshold $\varepsilon$ as well  (\autoref{eq:detection}). Consequently, in the presence of malicious control flows, \name's likelihood score $L$ did not exceed $\varepsilon$. Hence, the flows were classified as malicious correctly. This reduced \name's false negative and positive rates. 

\begin{table}[tp]
	\small
    \caption{Classification accuracy of all evaluation programs over four evaluation settings.}
	\begin{tabular}{  l | l  l  l  l }  
%     	\hline
		 Program & Time_HMM & Time_LSTM & Freq_HMM & \textbf{Freq_LSTM} \\
        \hline
        Matrix & 55\% &  52\% & 60\% & \textbf{100\%} \\
%         \hline
		Q-sort &  49\% & 60\%  & 41\% & \textbf{100\%} \\ 
%         \hline
        GD & 40\% &  64\%  & 40\% & \textbf{98\% } \\
%         \hline
        Newton & 48\% & 51\% & 63\% & \textbf{100\%} \\
%         \hline
        Conv & 57\% &  69\% & 56\% & \textbf{100\%} \\
%         \hline
        DCT & 53\% & 45\% & 51\% &  \textbf{94\%} \\
%         \hline
        Dijkstra & 62\% & 72\% & 65\% &  \textbf{100\%} \\
%         \hline
        AES & 50\% & 50\% & 67\% & \textbf{98\%} \\
%         \hline
        PID & 40\%  & 62\% & 71\% & \textbf{99\%} \\
%         \hline
        Patfilt & 51\% &  45\% & 67\% & \textbf{100\%} \\
%         \hline
	\end{tabular}   
    \label{tab:acc}
\end{table}

% AUC table
\begin{table}[tp]
	\small
    \caption{Area under curve (AUC) of all evaluated programs over all four evaluation settings.}
	\begin{tabular}{  l | l  l  l  l }  
%     	\hline
		 Program & Time_HMM & Time_LSTM & Freq_HMM & \textbf{Freq_LSTM} \\
        \hline
        Matrix & 0.34 &  0.52 & 0.90 & \textbf{0.99} \\
%         \hline
		Q-sort &  0.52 &  0.48 & 0.76 & \textbf{1.00} \\ 
%         \hline
        GD & 0.24 &  0.55  & 0.86 &  \textbf{0.98} \\
%         \hline
        Newton & 0.25 & 0.62 & 0.86 & \textbf{0.99} \\
%         \hline
        Conv & 0.62 & 0.65  & 0.81 & \textbf{0.99} \\
%         \hline
        DCT & 0.14 & 0.61 & 0.81 &  \textbf{0.99} \\
%         \hline
        Dijkstra & 0.44 & 0.51 & 0.85 &  \textbf{1.00} \\
%         \hline
        AES & 0.56 & 0.57 & 0.82 & \textbf{0.96} \\
%         \hline
        PID &  0.34 & 0.66 & 0.79 & \textbf{1.00} \\
%         \hline
        Patfilt & 0.22 & 0.73  & 0.87 & \textbf{0.99} \\
%         \hline
	\end{tabular}
    \label{tab:auc}
\end{table}

% likelihood score distribution results

% ROC curve

%Different thresholds need to be set on the likelihood scores under different scenarios, for example in applications where miss detection will cause huge damage, a lower threshold is desired to make sure most positive sample are included; however for those who want to avoid frequent false alarms, the threshold should be set higher. To evaluate the performance of our model on various thresholds, we compute the AUC. 
% As can be see from the figure, all programs have an ROC curve close to the rectangle, which represents perfect detection, their AUC are therefore all close to one. This shows that our model well separated positive and negative classes.

\paragraph{\textbf{Sliding window size.}} We investigated the influence of various sliding window sizes on \name accuracy. By using window of different sizes, \name essentially looks into the program execution at different granularities. A smaller window can capture finer grained transitions in the signal trace, but the frequency resolution of the spectra will be lower. This results in a less discriminative representation of the signal segments. Smaller windows also result in longer data sequence, therefore more recurrences of \name's neural network model. This makes the model less robust to random perturbations, since biases on the network states accumulate through the recurrences. A larger window size, on the other hand, will have spectra of better frequency resolution and better robustness, but some small transitions in the signal trace will be ignored. 

\begin{figure} [tp] 
\centering
\begin{subfigure}[b] {0.47\textwidth}
    \centering
    \includegraphics[width=\textwidth]{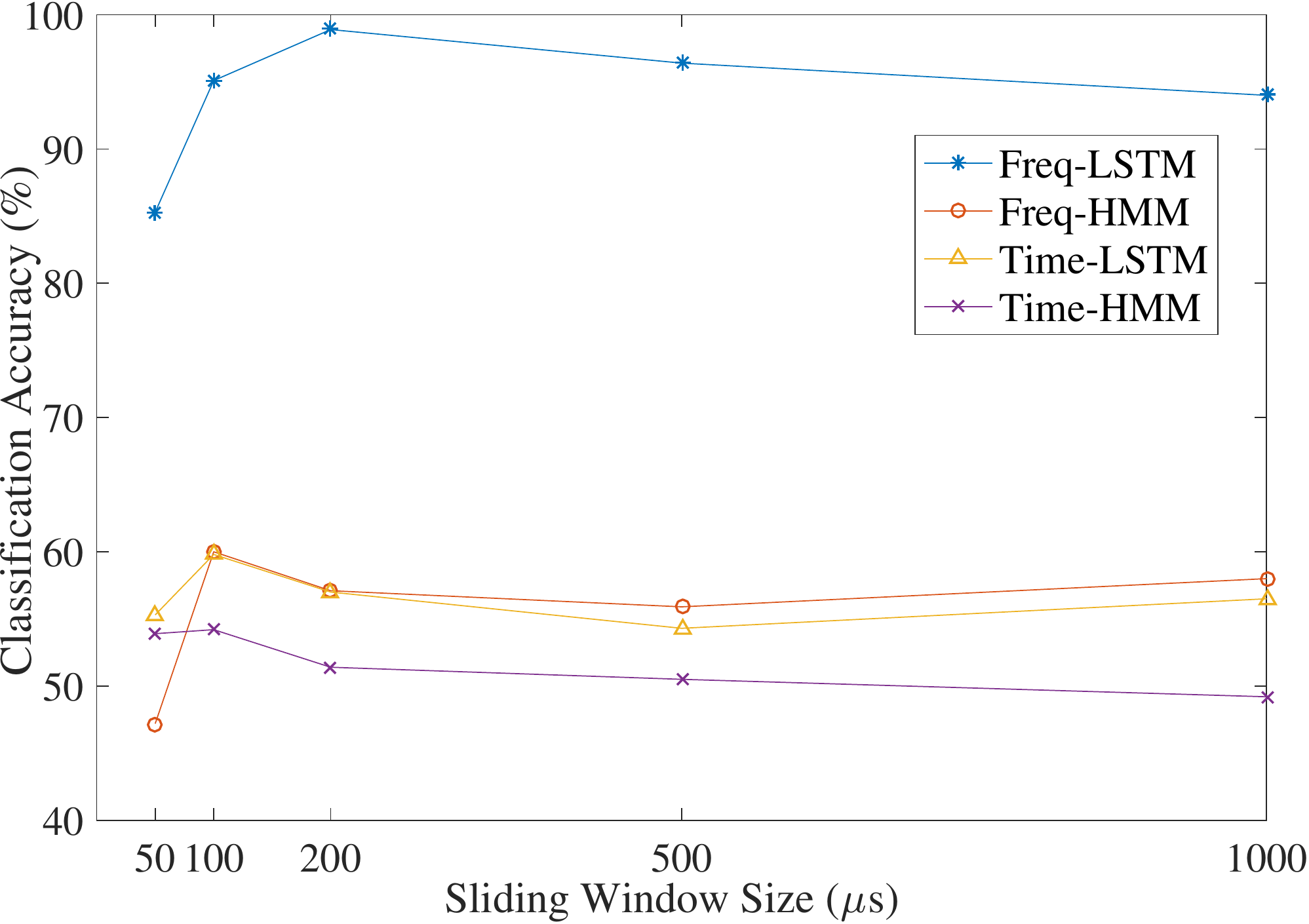}
   \caption{Classification accuracy vs. sliding window size.}
  \label{fig:acc}
\end{subfigure}

\vspace{0.15in}

\begin{subfigure}[b]{0.47\textwidth}
	\centering
	\includegraphics[width =\textwidth]{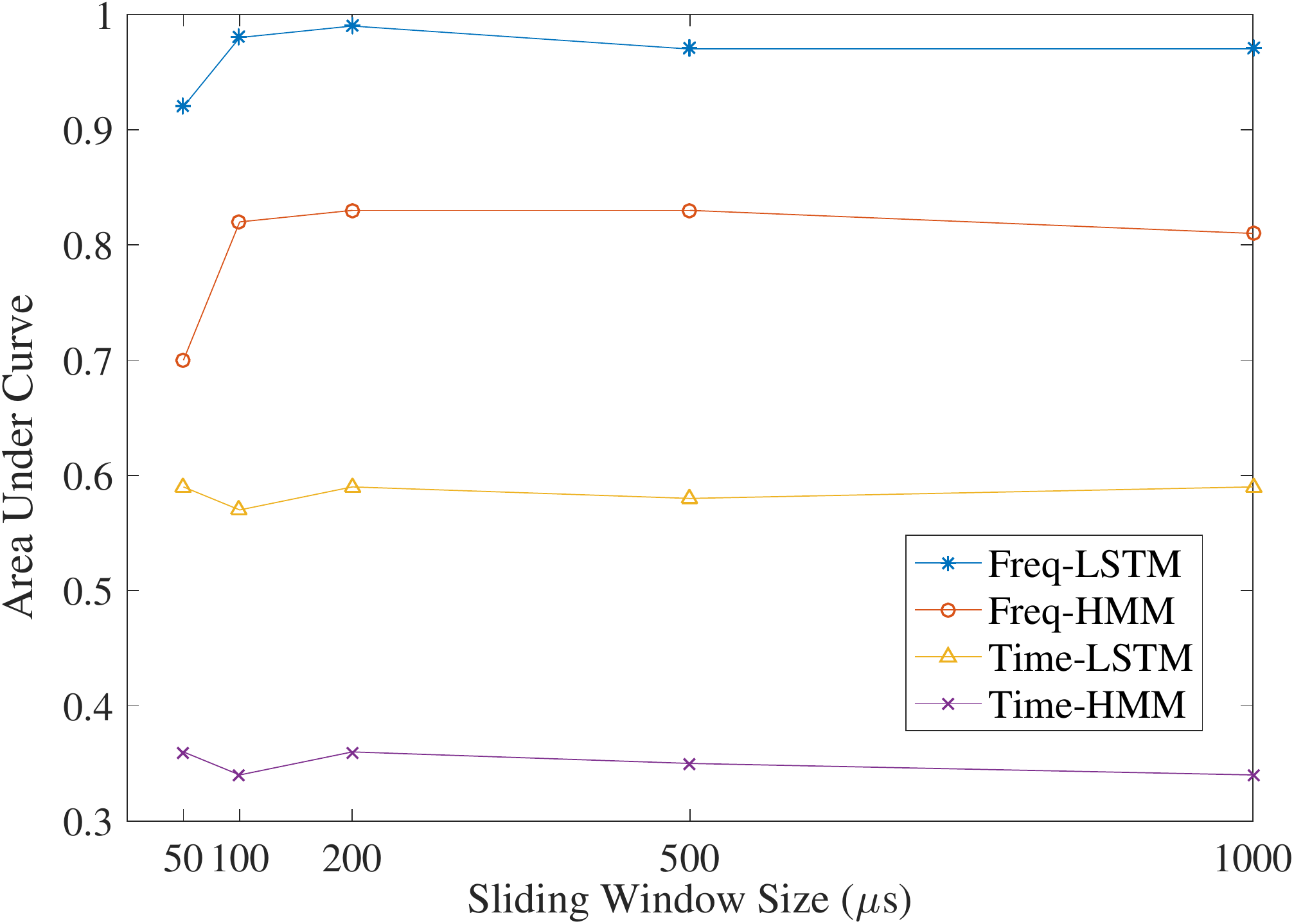}
 	\caption{Runtime vs. query trace length.}
	\label{fig:auc}
\end{subfigure}
\caption{Area under curve vs. sliding window size. }
\label{fig:windowsize}
\end{figure}

\autoref{fig:windowsize} shows how sliding window size affects \name accuracy (\autoref{fig:acc}) and AUC (\autoref{fig:auc}) both averaged on all ten applications. Using frequency data with a LSTM classifier outperforms all the other settings for all the window sizes. When the size of sliding window increases, both the classification and detection accuracy degrade because of the ignored useful transient information (as discussed above). Too small windows also cause accuracy degradation because of the resulting lower frequency resolution and less discriminative spectra. %This, as described above, is because larger window includes more instructions and compute a single spectrum, thus the important transient information and order information of these instructions are discarded when transferring signal traces into frequency representations. Also, smaller window produces spectra of lower frequency resolution. These spectra, as inputs to the model, are less discriminative. Therefore for too small window the performance degrades significantly. This trade off need to be carefully considered for specific application designs.  

\subsection{Performance}

We measured the time requirements to complete \name's classification of the collected EM traces. %A faster online classification (less process time) is desired since faster the malicious PLC programs are detected, less damage they will cause to the physical system. 
The required time includes the time of computing the spectrum sequence from the raw time domain EM signal trace if frequency representation is employed, and the time of passing the data sequence through the trained neural network model to get the prediction.

\begin{figure}[thbp] 
\centering
    \centering
    \includegraphics[width=.47\textwidth]{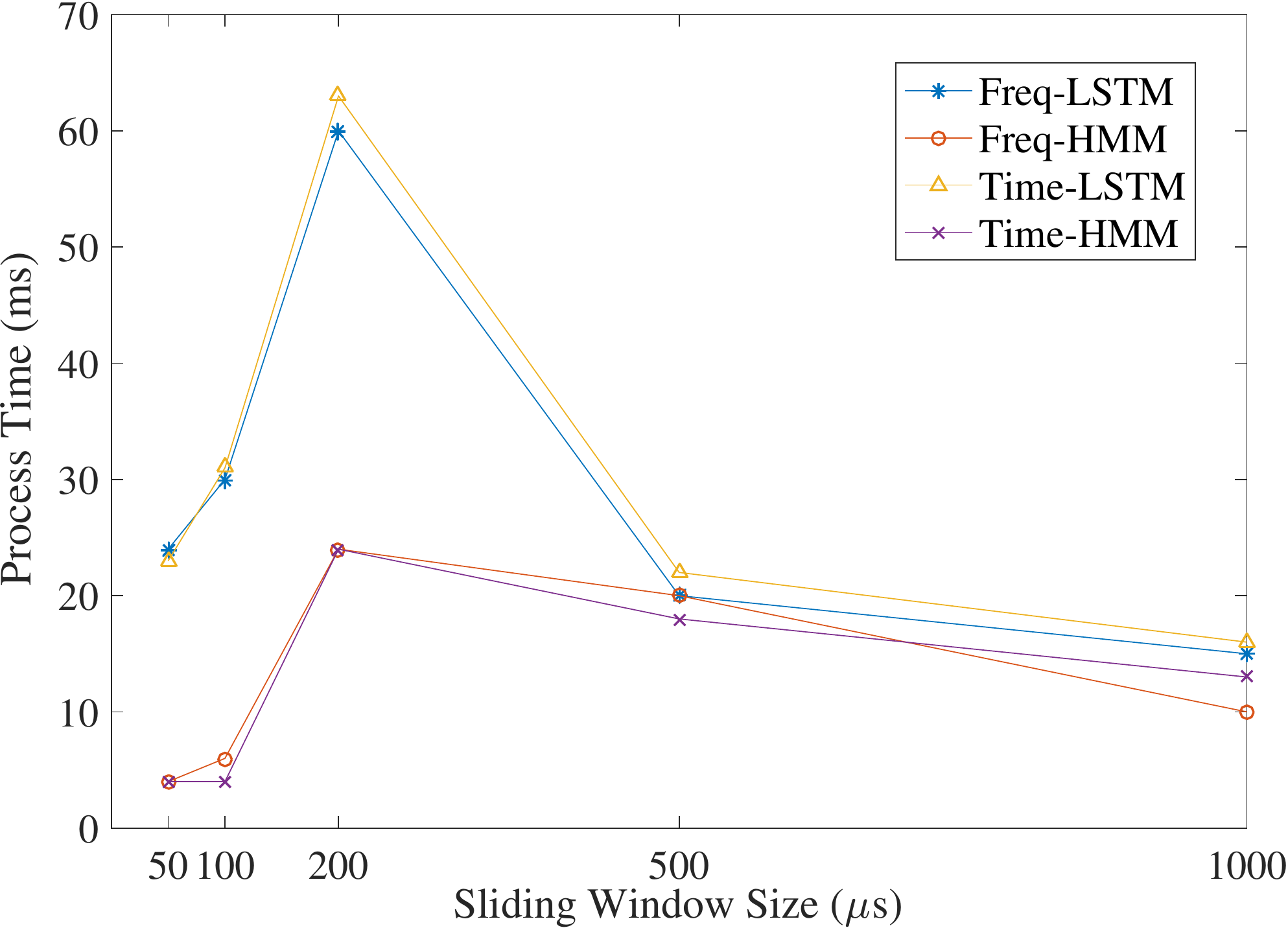}
\caption{Average process time of all the programs for the four evaluation settings.}
\label{fig:processtime}
\end{figure}

\autoref{fig:processtime} shows the average processing time for one input signal trace and various sliding window sizes. The numbers are averaged over all the ten applications. All the four evaluation settings are able to process the query signal within tens of milliseconds. LSTM-based approaches are overall slower than HMM-based solutions, because passing the input sequences through the network involves  more time-consuming array computations. HMM's faster speed comes at the cost of its remarkably lower classification and detection accuracy. 

The figure also shows that the larger sliding window sizes lead to reduced time requirements. This is expected as the larger sliding window produce shorter data sequences for a given EM signal trace. Consequently, there are fewer recurrences in the neural network. 

\section{Related Work}
\label{sec:related-work}

We discuss related work on controller and critical infrastructure security in terms of defense mechanisms and possible attacks. 

\paragraph{\textbf{Side channel analysis.}} There have been several recent works on analyzing side channels of various modalities for the purpose of  inspecting  runtime execution. On contactless monitoring, Eisenbarth et al.~\cite{eisenbarth2010building} determine the instruction types (not instances) by modeling individual instructions as HMM states. Msgna et al.~\cite{Msgna2013} perform similar analyses by modeling CFGs as HMMs. The authors assume equal-length basic blocks that is not often the case in practical settings. Other similar HMM-based program behavior modeling have also been studied~\cite{yeung2003host, warrender1999detecting, gao2009beyond, xu2015probabilistic}. On monitoring with contact requirements, cross correlation between the side channels (power~\cite{gonzalez2010detecting, gonzalez2009power} and RF traces~\cite{stone2013radio, stone2012radio, stone2015detecting}) and the program's single golden execution have been investigated for anomaly detection. However, obtaining a single golden execution is not feasible in practice. A complex PLC program may go through different execution paths depending on the inputs (sensor measurements). Vermon et al.~\cite{vermoen2007reverse} uses power signal side channels to reverse engineer the bytecode running on a Java smart card. Attacks to disclose substitution tables of the A3/A8 algorithm execution~\cite{novak2003side, clavier2004side} have been proposed. These methods focus on recovering the lookup table only. \name increases the accuracy of passive side-channel analysis of complete execution profiles using inexpensive contactless EM sensors. 

The most related work~\cite{liu2016code} provides code execution tracking based on the power signal, which  requires connections to an 11 $MHz$ 8-bit AVR microcontroller. The microprocessor is directly connected to the power supply using a single resistor. The sensor is a Tektronix MDO3034 oscilloscope with sampling rate of 1.25 $GHz$. \name provides \textit{contactless} execution monitoring of \textit{commercial} PLC processors (120 $MHz$ ARM Cortex M3 with three separate PCB boards for I/O, logic processing, and power supply) through a different \textit{modality} (electromagnetic emanations) and using \textit{lower-frequency} sensing sampling rates (10 $MHz$) with more than two orders of magnitude saving on the sensor cost. 

Another related work~\cite{nazari2017eddie} also performs execution monitoring and anomaly detection on IoT devices via the electromagnetic side channel. They looks at the prominent frequency peaks in the spectra of the signal segments as feature representations and models program executions with statistical distributions. \name uses the sequential neural network model to both extract discriminative features from signal segments and model the control flow transitions in a end to end manner. Moreover, they puts instrumentations at all the loop nests and examines them separately while \name looks at full executions without any instrumentation.

\paragraph{\textbf{Controller program analysis.}} Although a few processors contain a dedicated hardware unit for execution monitoring, e.g., embedded
trace macrocell~\cite{byrne2006shared}, many embedded controllers lack such hardware support. To analyze the software, offline control command verification solutions~\cite{mclaughlin2014tsv, park-chem-00, groote-95} implement formal methods to verify the safety of the control code immediately before it is executed on the PLC. They face scalability problem, caused by state-space explosion~\cite{mclaughlin2014tsv,sabot,huuck-05,biha-ieee-11}. %, it is almost always infeasible to explore all possible execution states of a given real controller program during a reasonable offline analysis time interval. 
Consequently, every control logic upload request by the operators, including the emergency cases, should wait for possibly minutes, hours, or more before the code is fully verified for PLC execution. Such delays are often unacceptable for real-time safety-critical control system operations. 

\paragraph{\textbf{Information security approaches.}} The related work to protect the control networks' trusted computing base (TCB) are insufficient as software patches are often applied only months after their  release~\cite{pollet-blackhat-10}, while new vulnerabilities are discovered on a regular basis~\cite{beresford-blackhat-11,basecamp}. The traditional perimeter-security tries to keep adversaries out of the protected control system entirely. Attempts include regulatory compliance approaches such as the NERC CIP requirements~\cite{nerc-cip} and access control~\cite{formby2016}. Despite the promise of information-security approaches, thirty years of precedence have shown the near impossibility of keeping adversaries out of critical systems~\cite{igure2006security} and less than promising results for the prospect of addressing the security problem from the perimeter~\cite{lewis2006critical, kuz2012construction, morris2009engineering}. Embedded controllers from most major vendors~\cite{kuz2012construction, valentine2013plc} and popular Human Machine Interfaces (HMIs)~\cite{morris2009engineering} have been shown to have fundamental security flaws.

%\noindent\textbf{Controller program analysis.} Basic static program analysis approaches use SAT-based model checking through Boolean logic~\cite{park-chem-00, groote-95, mclaughlin2012sabot} that could analyze sequence-based control systems with timers, but those are only narrowly applicable. Unlike \cite{canet-smc-00}, the two theorem-proving based approaches~\cite{huuck-05,biha-ieee-11} handle numerical instructions, but not do not implement rules for overflow checks or mixed bit vector and integer arithmetic. Almost all static analysis techniques \cite{mclaughlin2014tsv} fall short in either checking for all program details or scaling up to large-scale critical infrastructures. To improve  dynamic SCADA infrastructure monitoring techniques \cite{stouffer2011sp}, PLC-based approaches have been suggested \cite{berger2013automating, ioannides2004design} for dynamic physical plant monitoring. Dynamic plant behavior safety monitors~\cite{mohan-s3a-tr} and mathematical intrusion detectors~\cite{cheung-sri-07} are also related. In addition to being intrusive and causing performance overhead, dynamic monitoring solutions such as WeaselBoard~\cite{dhs-plc} focus mainly on accidental failures, ignoring malicious actions, and/or leave an insufficient time buffer for an effective response and recovery in case of an attack or failure. 

% \input{Sources/sec-discussions}
\section{Conclusions}
\label{sec:conclusions}

We presented \name, a contactless, passive, and non-intrusive control flow integrity monitoring solution for PLCs. \name identifies malicious code execution through side channel analyses of the controller's electromagnetic emanation signals. \name's data acquisition is done by an electromagnetic sensor, which provides an air gap between the trusted computing bases of the target PLC and \name's monitoring engine. Our empirical studies with commercial PLC controllers and several real application binaries show \name can monitor high frequency commercial processors with low frequency sensor sampling. \name can detect malicious code executions on popular Allen Bradley PLCs with $\%98.9$ accuracy and with zero runtime overhead by its design.

\section*{Acknowledgement}

We would like to thank the National Science Foundation (NSF) for supporting this research project.

\bibliographystyle{ACM-Reference-Format}
\bibliography{refs}

\end{document}